\documentclass[11pt,onecolumn]{article}
\usepackage[english]{babel}

\usepackage{amsmath}
\usepackage{amsfonts}
\usepackage{amssymb}
\usepackage{graphicx}
\usepackage[affil-it]{authblk}
\usepackage{hyperref}
\usepackage{color}
\usepackage{caption}
\usepackage{verbatim}
\usepackage[normalem]{ulem}
\usepackage{subfigure}
\usepackage{slashed}
\usepackage{cancel}
\usepackage{braket}
\usepackage{multicol}
\usepackage{tcolorbox}
\tcbuselibrary{theorems}
\usepackage{bm}
\usepackage{enumitem}
\usepackage[numbers,sort&compress]{natbib}

\hypersetup{linktocpage=true}

\usepackage{tikz} 
\usepackage{tikz-feynman}

\tikzfeynmanset{compat=1.0.0}

\usetikzlibrary{arrows,shapes}
\usetikzlibrary{trees}
\usetikzlibrary{matrix,arrows} 				
\usetikzlibrary{positioning}				
\usetikzlibrary{calc,through}				
\usetikzlibrary{decorations.pathreplacing}  
\usepackage{pgffor}							

\usetikzlibrary{decorations.pathmorphing}	
\usetikzlibrary{decorations.markings}
\tikzset{
    vector/.style={decorate, decoration={snake}, draw},
	provector/.style={decorate, decoration={snake,amplitude=2.5pt}, draw},
	antivector/.style={decorate, decoration={snake,amplitude=-2.5pt}, draw},
    fermion/.style={draw=black, postaction={decorate},
        decoration={markings,mark=at position .55 with {\arrow[draw=black]{>}}}},
    fermionbar/.style={draw=black, postaction={decorate},
        decoration={markings,mark=at position .55 with {\arrow[draw=black]{<}}}},
    fermionnoarrow/.style={draw=black},
    gluon/.style={decorate, draw=black,
        decoration={coil,amplitude=4pt, segment length=5pt}},
    scalar/.style={dashed,draw=black, postaction={decorate},
        decoration={markings,mark=at position .55 with {\arrow[draw=black]{>}}}},
    scalarbar/.style={dashed,draw=black, postaction={decorate},
        decoration={markings,mark=at position .55 with {\arrow[draw=black]{<}}}},
    scalarnoarrow/.style={dashed,draw=black},
    electron/.style={draw=black, postaction={decorate},
        decoration={markings,mark=at position .55 with {\arrow[draw=black]{>}}}},
	bigvector/.style={decorate, decoration={snake,amplitude=4pt}, draw},
    line/.style={draw=black},
}\usetikzlibrary{decorations.markings}

\usepackage{fancyhdr}
\fancypagestyle{plain}{%
	\fancyhead[R]{\Large{TUM-HEP 1338/20}}
	
}

\title{Fermion Singlet Dark Matter in a Pseudoscalar Dark Matter Portal}
\author[1]{Basti\'an D\'iaz S\'aez\thanks{bastian.diaz@tum.de}}
\author[2]{Patricio Escalona\thanks{patricio.escalona@sansano.usm.cl}}
\author[3]{Sebasti\'an Norero\thanks{sebastian.norero@puc.cl}}
\author[4]{Alfonso Zerwekh\thanks{alfonso.zerwekh@usm.cl}}
\affil[1]{Physik-Department, Technische Universität München, James-Franck-Straße, 85748 Garching, Germany}
\affil[2]{Departamento de Física, Universidad Técnica Federico Santa María, Avenida España 1680, Valparaíso, Chile}
\affil[3]{Instituto de Física, Pontificia Universidad Católica de Chile, Avenida Vicuña Mackenna 4860, Santiago, Chile}
\affil[4]{Departamento  de  Física y Centro  Científico-Tecnológico de Valparaíso, Universidad Técnica Federico Santa María, Avenida España 1680, Valparaíso, Chile}

\begin{document}
\maketitle


\begin{abstract}
We explore a simple extension to the Standard Model containing two gauge singlets: a Dirac fermion and a real pseudoscalar. In some regions of the parameter space both singlets are stable without the necessity of additional symmetries, then becoming a possible two-component dark matter model. We study the relic abundance production via freeze-out, with the latter determined by annihilations, conversions and semi-annihilations. Experimental constraints from invisible Higgs decay, dark matter relic abundance and direct/indirect detection are studied. We found three viable regions of the parameter space, and the model is sensitive to indirect searches.
\end{abstract}

\newpage
\tableofcontents


\section{Introduction}
Astrophysical evidence of dark matter (DM) has been accumulating for more than forty years now, but its fundamental nature remains unknown. From the particle physics points of view, different approaches have been carried out over the years to account for the elusive DM (for a review see \cite{RevModPhys.90.045002}), and in particular, simple extensions to the SM containing gauge singlets look appealing for their simplicity, DM predictions and testable phenomenology \cite{Silveira:1985rk, McDonald:1993ex, Burgess:2000yq, Cline:2013gha, Kim:2008pp, Escudero:2016gzx, GAMBIT:2017gge}. Nowadays, those WIMP minimal extensions have been very constrained, especially by direct detection, motivating other possible alternatives. For instance, the interplay of a (pseudo)scalar and a fermion, both gauge singlets, open up the possibilities in many aspects: multi-component DM, new interaction channels, novel experimental signatures, small-scale structures, among others \cite{Pospelov:2007mp, Cao:2007fy,Zurek:2008qg, Baek:2011aa, LopezHonorez:2012kv, Heikinheimo:2013xua, Ghorbani:2014qpa, Kim:2016csm,Bhattacharya:2013hva, Esch:2013rta, Esch:2014jpa, Ipek:2014gua, Cai:2015zza, Konig:2016dzg, Ahmed:2017dbb, Kahlhoefer:2017umn, Baek:2017vzd, Arcadi:2017wqi, Ghorbani:2018pjh, Bhattacharya:2018cgx, Duch:2019vjg}.

Keeping minimality, in this work we study a gauge singlet sector composed of a real pseudoscalar and a Dirac fermion. Depending on the coupling values and mass hierarchy between the singlets, the model admits a variety of DM productions with either one or the two singlets being stable. Two new interactions are present in the model: a Higgs portal and a dark sector coupling. The Higgs interaction is key because it regulates to what extent the dark sector is coupled to the SM, whereas the internal dark sector coupling only regulates the coupling between the two singlets. In our knowledge, we study for the first time the WIMP regime of this framework in which both couplings take sizable values such that both singlets were in thermal equilibrium with the SM bath in the early universe. Interestingly, in certain mass hierarchy between the two singlets, the stability of both fields is guaranteed by a parity symmetry without the necessity of introducing new ad hoc discrete symmetries. Models with this last feature or accidental symmetries have been studied in different DM context, such as \textit{Minimal Dark Matter} \cite{Cirelli:2005uq}, spontaneous symmetry breaking \cite{Walker:2009en}, two DM components \cite{Bernal:2018aon}, vector DM \cite{Saez:2018off, Belyaev:2018xpf} and rank-two fields \cite{Cata:2014sta}. 

In the model under consideration, the DM relic abundance is triggered by annihilations, DM conversions \cite{Belanger:2011ww} and semi-annihilations \cite{DEramo:2010keq, Belanger:2012vp}, showing remarkable features in some regions of the parameter space. Further, we constraint the model considering the measured relic abundance in the universe, Higgs invisible decay and direct/indirect detection. For the latter, we explore box-shaped gamma ray spectra \cite{Nomura:2008ru, Ibarra:2015tya}, and we confront the available parameter space with Fermi-LAT data, CTA projections and AMS-02 bounds. 

The paper is organized as follows. In section 2 we present the model and its theoretical constraints. In section 3 we explore the possible DM relic abundance mechanisms presents in the model, with a precise analysis of the two-component freeze-out scenario. In section 4 we review experimental constraints and the available parameter space along with indirect detection signals. In the last section we discuss and state our conclusions. 

\section{The model}
The model adds to the SM two gauge singlets: one Dirac fermion $\psi$ and a pseudo-scalar $s$. Under a parity transformation, the fields transform as $\psi \rightarrow \gamma^0\psi$ and $s\rightarrow -s$, giving rise to the following Lagrangian: 
\begin{eqnarray}\label{lag1}
 \mathcal{L} = \mathcal{L}_{SM} + \bar{\psi}(i\slashed{\partial} - m_\psi)\psi + \frac{1}{2}(\partial_\mu s)^2 + ig_{\psi}s\bar{\psi}\gamma_5\psi - V(H , s),
\end{eqnarray}
where the scalar potential is given by
\begin{eqnarray}
 V(H , s) &=& \mu^2\vert H \vert^2 + \lambda_H \vert H \vert^4 + \frac{\mu_s^2}{2}s^2 + \frac{\lambda_s}{4!}s^4 + \frac{\lambda_{hs}}{2}\vert H \vert^2 s^2,
\end{eqnarray}
with $H$ being the Higgs doublet\footnote{One may consider $\psi$ as a sterile neutrino that mixes with the active ones. If the mixing is small enough, the sterile neutrino may be stable on cosmological scales and can be produced through active-sterile oscillations. In this work we assume that the mixing angle is sufficiently small to avoid these effects. For a discussion in this direction see \cite{Merle:2015vzu, Konig:2016dzg, Heikinheimo:2016yds}.}. We make $g_\psi$ real under a fermion field redefinition such that the theory is CP-conserving. We assume that the singlet scalar does not acquire vacuum expectation value (vev), and after EWSB in the unitary gauge, $H = (0, (v_H + h)/\sqrt{2})^T$ with $v_H = 246$ GeV the Higgs vev, the scalar potential may be rewritten as 
\begin{eqnarray}
 V(h , s) &=&  \frac{\mu_s^2}{2}s^2 + \frac{\lambda_s}{4!}s^4 + \frac{\lambda_{hs}}{4}(h^2 + 2v_Hh) s^2,
\end{eqnarray}
with the mass of the scalars given by 
\begin{eqnarray}
 m_h^2 &=& 2v_H^2 \lambda_H, \\
 m_s^2 &=& \mu_s^2 + \lambda_{hs}v_H^2/2.
\end{eqnarray}
Here we consider $m_h = 125$ GeV. The stability of the fermion is easily recognized due to the fact that it appears in pairs in the Lagrangian. In the scalar sector $s$ appears in pairs, forbidding its decay. The linear term in $s$ in \ref{lag1}, imply that as $m_s \geq 2m_\psi$ the pseudoscalar may decay into a pair of $\psi$, whereas as $m_s < 2m_\psi$ the scalar singlet becomes stable at tree level and at all orders in perturbation theory. In the following we argument the latter fact.
\begin{figure}[t!]
\centering
\includegraphics[width=0.55\textwidth]{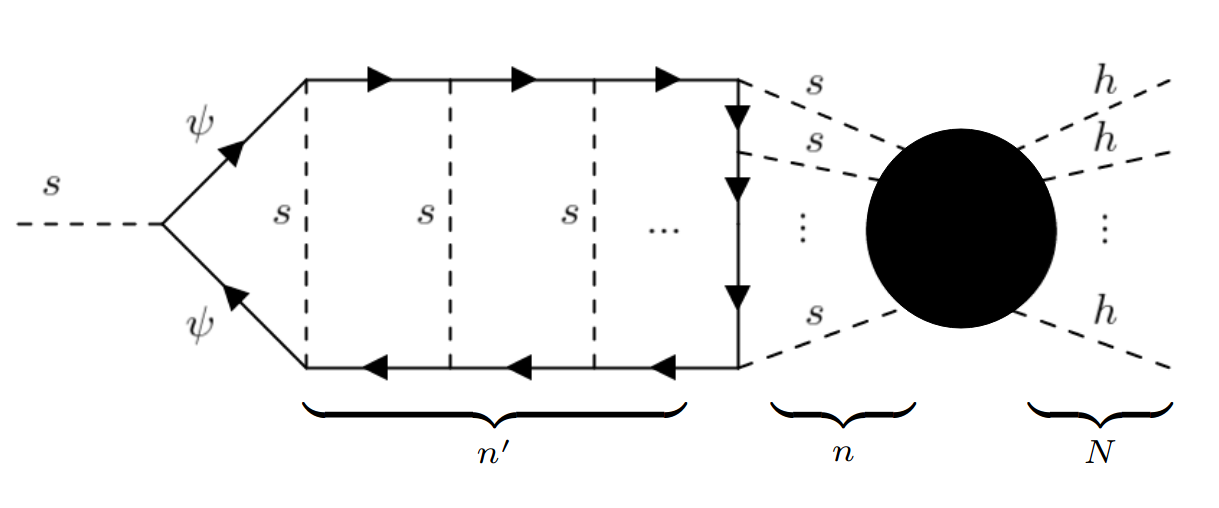}
\caption{\textit{General radiative $s$ decay. The values for $n'$ refers to internal $s$ lines in the closed fermionic loop, $n$ for outgoing $s$ lines and $N$ for Higgs lines. The black globe takes into account arbitrary interactions among $s$ and $h$ fields.}}
\label{loop}
\end{figure}

The possible decay of $s$ at an arbitrary number of loops is represented in Fig.~\ref{loop}, with the decay of $s$ followed by a singlet fermion closed loop and the black circle representing possible interactions between an arbitrary number of $s$ and $h$. In the figure $n', n$ and $N$ simply represent the number of scalar lines. For simplicity, let us start assuming $n' = 0$. If $n$ is even, the resulting fermion trace will at most contain terms of the form $\epsilon^{\mu \nu \rho \dots}p_\mu p_ \nu p_\rho\dots$, which vanishes exactly. If $n$ is odd, the trace is different than zero, but there is no way to connect an odd number of outgoing $s$ with an arbitrary number of $N$ Higgs boson in the black bloop, due to the presence of the CP symmetry of the scalar potential. Now, if $n'\neq 0$, the previous arguments still remains, because internal lines of $s$ in the fermion loop only add an even number of both $\gamma^5$ and fermion propagators to the trace, only adding vanishing contributions without changing the final result. In consequence, from a perturbative point of view, the stability is guaranteed for the pseudoscalar singlet \footnote{One can introduce an axion-like anomalous 5-dimensional effective operator that conserves CP \cite{Mambrini:2015sia}, $\mathcal{L}\supset \frac{\lambda}{\Lambda} s G_{\mu \nu}  \Tilde{G}^{\mu \nu}$, with $\lambda$ a dimensionless coupling, $\Lambda$ some high energy scale, $G$ the field strength of any gauge field and $\Tilde{G}$ its dual. This operator induce the decay of the pseudoscalar singlet into gauge bosons. For instance, considering $m_s =  10^3 \: \mathrm{GeV}$, we require $\lambda \lesssim 10^{-7}$ for a Planck scale induced operator in order to have a cosmologically stable particle. For a GUT scale induced operator we require $\lambda \lesssim 10^{-10}$. 
Other possible low energy origin of such operator requires the introduction of heavy vector-like fermions \cite{Lee:2012ph}, which are not part of our model construction.}.

Finally, theoretical constraints put bounds on the free parameters of the model. The stability of the electroweak vacuum for $s$ and $h$ imposes that \cite{Lerner:2009xg}
\begin{eqnarray}
 \lambda_s > 0,\quad \lambda_{hs} > -\sqrt{\frac{2}{3}\lambda_H\lambda_s} ,
\end{eqnarray} 
with $\lambda_{H} = m_h^2/(2v_H^2) \simeq 0.1$. From perturbativity we set that $|g_\psi|, |\lambda_s| < 4\pi$ to ensure that loop corrections are smaller than tree-level processes \cite{Lerner:2009xg}. Unitarity constraints are less stringent than the upper limits based on our perturbativity criteria \cite{Cynolter:2004cq, Cline:2013gha}. The signs of $g_\psi$ and $\lambda_{hs}$ are not relevant for the analysis in this work due to the fact that the relevant processes depend quadratically on them, therefore the theoretical constraints set $0 <\lambda_{hs} < 4\pi$ and $0 <g_\psi < 4\pi$. 

\begin{figure}[t!]
\centering
\includegraphics[width=0.34\textwidth]{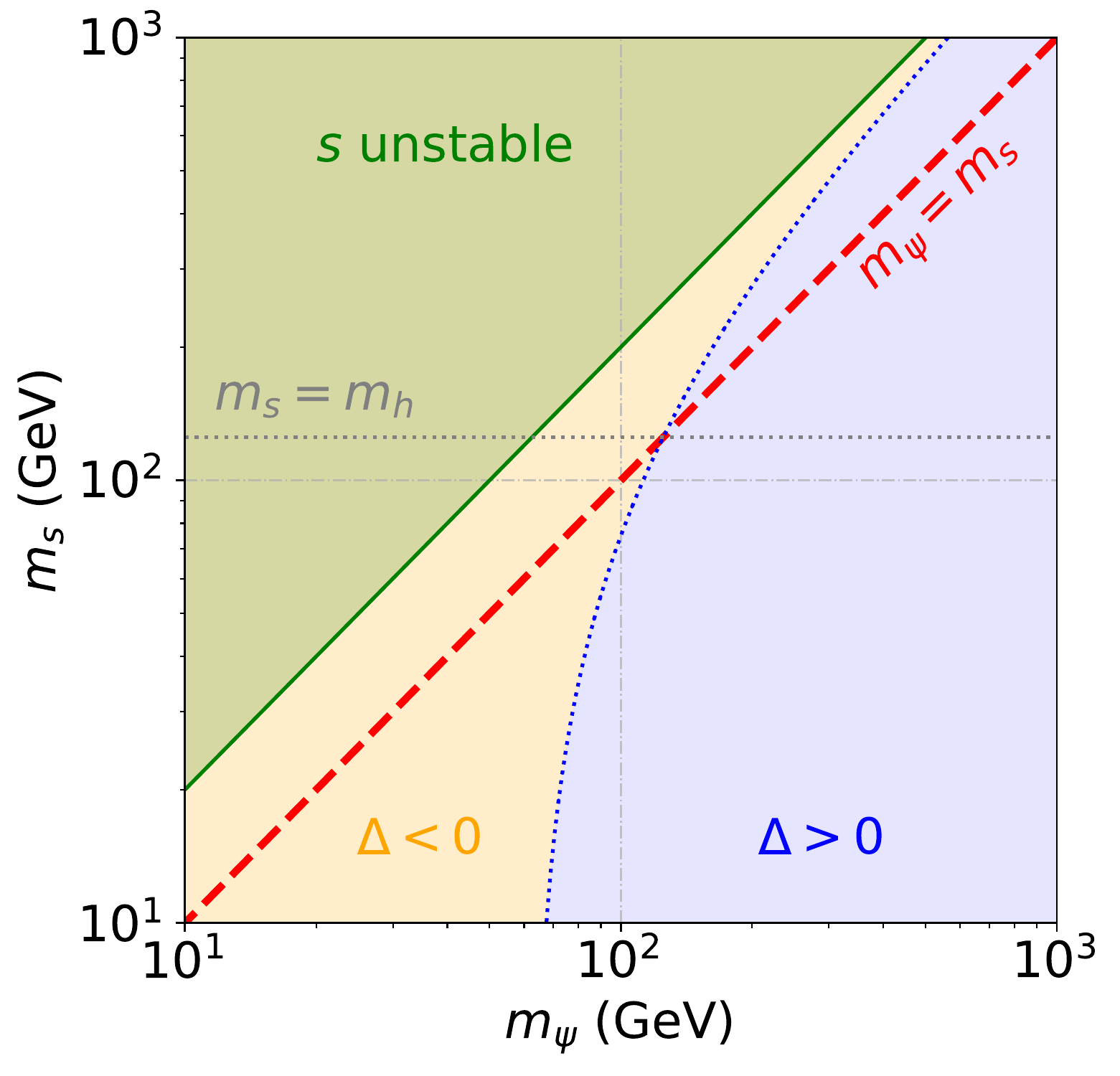}
\caption{\textit{Mass plane, with the green region indicating $s$ unstable, whereas in the orange and blue regions both singlets are stable. The blue one indicates where $s$-channel semi-annihilations are present, with $\Delta \equiv 2m_\psi - m_s - m_h \geq 0$ (for details see Sec.~\ref{2dmsec}).}}
\label{parameter}
\end{figure}

\section{Relic Density}
\subsection{Parameter Space}
Depending on the intensity of the couplings and the mass hierarchy between $m_\psi$ and $m_s$, the model may present different DM scenarios with one or two stable particles. When $\lambda_{hs}$ takes very small values, i.e. $\lambda_{hs}\sim 10^{-12} - 10^{-6}$, the singlet sector never thermalize with the SM, then the DM production may occurs via freeze-in and/or dark freeze-out \cite{Heikinheimo:2016yds, Kainulainen:2016vzv}. On the contrary, as the Higgs portal coupling takes sizable values, the singlet scalar enters into thermal equilibrium with the SM (this work). Based on the latter fact, two possible coupling regimes concerning $g_\psi$ may be present:
\begin{itemize}
 \item $g_\psi \lesssim 10^{-6}$: The two singlets in the dark sector will interact feebly. If $m_s > 2m_\psi$, $s$ becomes unstable, and $\psi$ may be produced via freeze-in through the decay of $s$ and $2\rightarrow 2$ scattering processes. The green region in Fig.~\ref{parameter}~ shows this parameter space in terms of the mass hierarchy. For $m_s \leq 2m_\psi$, $s$ becomes stable and  define a DM candidate identical to that of the \textit{Singlet Higgs Portal} model (SHP) \cite{Cline:2013gha, GAMBIT:2017gge}, since $\psi$ does not interfere in the dynamic of the former due to their feebly interactions. The SHP model has been exhaustively studied previously, and displays a highly constrained parameter space around the EW scale. 
 \item $g_\psi \gtrsim 10^{-6}$: $s$ will bring $\psi$ fast into the thermal equilibrium. From the relic abundance point of view, the only relevant case here is when both singlets are stable, i.e. $m_s < 2m_\psi$ (orange and blue region in Fig.~\ref{parameter}~), otherwise $\psi$ would not have any channel to annihilate, giving rise to an overabundance (see Feynman diagrams in Fig.~\ref{diagrams}). Based on the latter point, different type of interactions appear that determine the relic abundance of each singlet.
\end{itemize}
In this work, we focus on this last scenario, in which both singlets are stable. It is worth to mention that even when the first case with one DM candidate via freeze-in is a perfectly viable DM model, it presents a challenge phenomenology \cite{Blennow:2013jba} (see however \cite{Belanger:2018sti}). Recently, novel scenarios with \textit{inverse semi-production} have been proposed with interesting dark matter production and phenomenology \cite{Bringmann:2021lyf, Hryczuk:2021qtz}. 

\subsection{Boltzmann equation}\label{2dmsec}
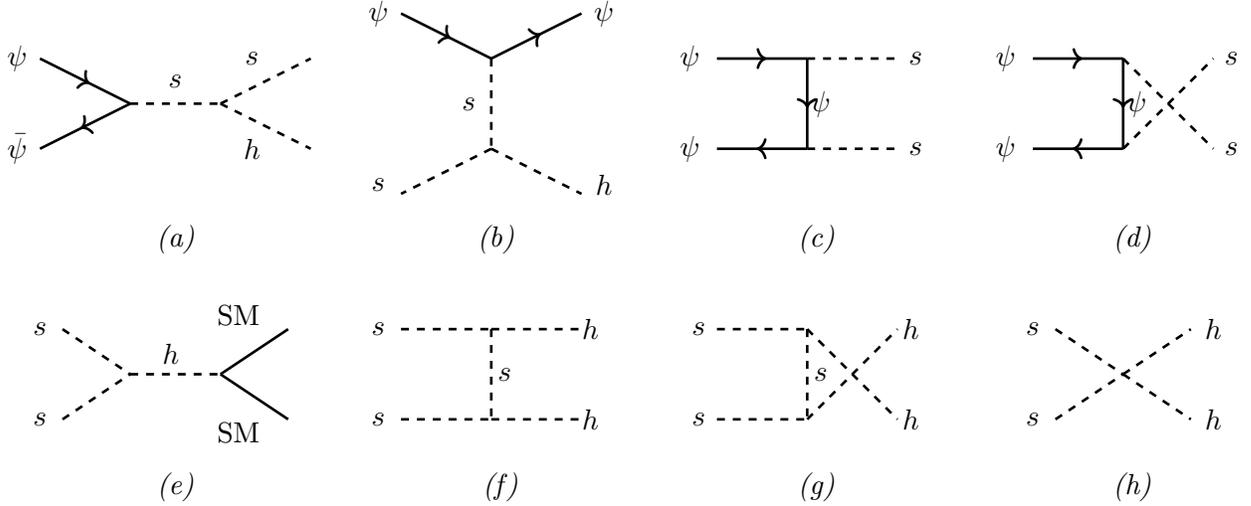
\begin{figure}[t!]
\centering
\begin{tikzpicture}[line width=1.0 pt, scale=0.6]
\begin{scope}[shift={(0,5)}]
	\draw[fermion](-3,1) -- (-1,0);
	\draw[fermionbar](-3,-1) -- (-1,0);
	\draw[scalarnoarrow](-1,0) -- (1,0);
	\draw[scalarnoarrow](1,0) -- (3,1);
	\draw[scalarnoarrow](1,0) -- (3,-1);
   \node at (-3.5,1.0) {$\psi$};
	\node at (-3.5,-1.0) {$\bar{\psi}$};
   \node at (0,0.5) {$s$};
	\node at (1.7,1) {$s$};
   \node at (1.7,-1) {$h$};
   
     \node at (0,-3) {$\textit{(a)}$};
\end{scope}

\begin{scope}[shift={(8,6)}]
	\draw[fermion](-3,1) -- (-1,0);
	\draw[fermion](-1,0) -- (1,1);
	\draw[scalarnoarrow](-1,0) -- (-1,-2);
	\draw[scalarnoarrow](-1,-2) -- (-3,-3);
	\draw[scalarnoarrow](-1,-2) -- (1,-3);
   \node at (-3.5,1.0) {$\psi$};
	\node at (1.5,1.0) {$\psi$};
   \node at (-1.5,-1) {$s$};
	\node at (-3.5,-2.8) {$s$};
   \node at (1.5,-2.8) {$h$};
        \node at (-0.9,-4) {$\textit{(b)}$};
\end{scope}

\begin{scope}[shift={(15,5)}]
 \draw[fermion](-3,1) -- (-1,1);
	\draw[fermionbar](-3,-1) -- (-1,-1);
	\draw[fermion](-1,1) -- (-1,-1);
	\draw[scalarnoarrow](-1,1) -- (1,1);
	\draw[scalarnoarrow](-1,-1) -- (1,-1);
    \node at (-3.6,1.0) {$\psi$};
	\node at (-3.6,-1.0) {$\psi$};
    \node at (-0.7,0) {$\psi$};
	\node at (1.4,1) {$s$};
    \node at (1.4,-1) {$s$};
     
    \node at (-0.8,-3) {$\textit{(c)}$};
\end{scope}

\begin{scope}[shift={(22,5)}]
	\draw[fermion](-3,1) -- (-1,1);
	\draw[fermionbar](-3,-1) -- (-1,-1);
	\draw[fermion](-1,1) -- (-1,-1);
	\draw[scalarnoarrow](-1,1) -- (1,-1);
	\draw[scalarnoarrow](-1,-1) -- (1,1);
    \node at (-3.6,1.0) {$\psi$};
	\node at (-3.6,-1.0) {$\psi$};
    \node at (-0.7,0) {$\psi$};
	\node at (1.4,1) {$s$};
    \node at (1.4,-1) {$s$};
    
    \node at (-0.8,-3) {$\textit{(d)}$};
\end{scope}

\begin{scope}[shift={(0,-1)}]
	\draw[scalarnoarrow](-2.5,1) -- (-1,0);
	\draw[scalarnoarrow](-2.5,-1) -- (-1,0);
	\draw[scalarnoarrow](-1,0) -- (1,0);
	\draw[line](1,0) -- (2.5,1);
	\draw[line](1,0) -- (2.5,-1);
    \node at (-3,1.0) {$s$};
	\node at (-3,-1.0) {$s$};
    \node at (-0.1,0.46) {$h$};
	\node at (1.4,1.3) {SM};
    \node at (1.4,-1.3) {SM};
    
    \node at (-0,-2.5) {$\textit{(e)}$};
\end{scope}

\begin{scope}[shift={(8,-1)}]
	\draw[scalarnoarrow](-3,1) -- (-1,1);
	\draw[scalarnoarrow](-3,-1) -- (-1,-1);
	\draw[scalarnoarrow](-1,1) -- (-1,-1);
	\draw[scalarnoarrow](-1,1) -- (1,1);
	\draw[scalarnoarrow](-1,-1) -- (1,-1);
    \node at (-3.5,1.0) {$s$};
	\node at (-3.5,-1.0) {$s$};
    \node at (-0.7,0) {$s$};
	\node at (1.2,1) {$h$};
    \node at (1.2,-1) {$h$};
        
     \node at (-0.8,-2.5) {$\textit{(f)}$};
\end{scope}

\begin{scope}[shift={(15,-1)}]
	\draw[scalarnoarrow](-3,1) -- (-1,1);
	\draw[scalarnoarrow](-3,-1) -- (-1,-1);
	\draw[scalarnoarrow](-1,1) -- (-1,-1);
	\draw[scalarnoarrow](-1,1) -- (1,-1);
	\draw[scalarnoarrow](-1,-1) -- (1,1);
    \node at (-3.4,1.0) {$s$};
	\node at (-3.4,-1.0) {$s$};
    \node at (-0.7,0) {$s$};
	\node at (1.3,1) {$h$};
    \node at (1.3,-1) {$h$};
    
        \node at (-0.8,-2.5) {$\textit{(g)}$};
\end{scope}

\begin{scope}[shift={(22,-1)}]
	\draw[scalarnoarrow](-2.5,1) -- (-1,0);
	\draw[scalarnoarrow](-2.5,-1) -- (-1,0);
	\draw[scalarnoarrow](-1,0) -- (0.5,1);
	\draw[scalarnoarrow](-1,0) -- (0.5,-1);
    \node at (-3,1.0) {$s$};
	\node at (-3,-1.0) {$s$};
	\node at (1,1) {$h$};
    \node at (1,-1) {$h$};
            
    \node at (-0.8,-2.5) {$\textit{(h)}$};
\end{scope}
\end{tikzpicture}
\caption{\textit{Relevant processes at the freeze-out time in the two DM component model. Diagrams (\textit{a}) and  (\textit{b}) are the $s$- and $t$-channel semi-annihilations, respectively, (\textit{c}) and (\textit{d}) correspond to DM conversions, and (\textit{e}) to (\textit{h}) are the annihilations of the pseudoscalar into SM and Higgs particles.}}\label{diagrams}
\end{figure}
As we pointed out before, the two-component scenario of our interest occurs for $m_s < 2m_\psi$ and when the couplings $(g_\psi , \lambda_{hs})$ are sizable. In this case, at temperatures higher than the individual masses of the singlet, both DM components are in thermal equilibrium with the SM. The departure of the equilibrium occurs once the temperature goes below the masses of the singlets, and three types of scattering participate in this process: annihilations, semi-annihilations, and dark matter conversions (Fig.~\ref{diagrams}). Based on \cite{Belanger:2014vza}, the evolution of the individual singlet abundances $Y_i \equiv n_i/s$, with $i=\psi , s $, as a function of the temperature $x \equiv \mu /T$, with $\mu = m_\psi m_s /(m_\psi + m_s)$, is given by 
\begin{eqnarray}\label{boltzc}
 \frac{dY_\psi}{dx} &=& - \lambda_{\psi\bar{\psi}ss}\left(Y_\psi^2 - Y_s^2\frac{Y_{\psi ,e}^2}{Y_{s,e}^2}\right) - \lambda_{\psi\bar{\psi}sh}\left(Y_\psi^2 - Y_s\frac{Y_{\psi ,e}^2}{Y_{s,e}}\right), \nonumber \\
 \frac{dY_s}{dx} &=& - \lambda_{ssXX} \left(Y_s^2 - Y_{s,e}^2\right) + \lambda_{\psi\bar{\psi}ss} \left(Y_\psi^2 - Y_s^2\frac{Y_{\psi ,e}^2}{Y_{s,e}^2}\right) \nonumber\\ 
                      & + & \frac{1}{2}\lambda_{\psi\bar{\psi}sh}\left(Y_\psi^2 - Y_s\frac{Y_{\psi ,e}^2}{Y_{s,e}}\right) -\frac{1}{2} \lambda_{s\psi\psi h}\left(Y_sY_\psi - Y_\psi Y_{s,e}\right),
\end{eqnarray}
where we have defined
\begin{align}
	\lambda_{ijkl}(x):=
	\dfrac{\langle\sigma_{ijkl}v\rangle(x)\cdot s(T)}
	{x\cdot H(T)}, \qquad \text{for} ~ i,j,k,l=\psi,s,h,X.
	\label{def_lambda}
	\end{align}
with $X$ referring to a SM particle, $\braket{\sigma v}$ the thermally averaged cross section, and the entropy density $s$ and Hubble rate $H$ in a radiation dominated universe given by
	\begin{align}\label{Hubbleandentropy}
	H(T)=\sqrt{\dfrac{4\pi^3G}{45}g_{*}(T)} \cdot T^2, \quad 
	s(T)=\dfrac{2\pi^2}{45}g_{*s}(T)\cdot T^3,
	\end{align}
where $G$ is the Newton gravitational constant, and $g_*$ and $g_{s*}$ are the effective degrees of freedom contributing respectively to the energy and the entropy density at temperature $T$ \cite{Husdal:2016haj}. The equilibrium densities, $Y_{i,e} \equiv n_{i,e}/s$, are calculated using the Maxwell-Boltzmann distribution, whose number density is given by
	\begin{align}\label{eqden}
	n_{ie}(T) = g_i\dfrac{m_i^2}{2\pi^2}TK_2(\tfrac{m_i}{T}),
	\end{align}
with $g_i$ the internal spin degrees of freedom, and $K_2$ is the modified Bessel function of the second kind. The DM density parameter is given by
\begin{align}\label{relic}
	\Omega h^2 = \sum_{i=\psi, s}\Omega_i h^2, ~~\qquad
	\Omega_i h^2 = \dfrac{2.9713\cdot 10^9}{10.5115 ~\text{GeV}}
	\cdot Y_{i,0} \cdot m_i ,
	\end{align}
where $Y_{i,0}$ is the yield of each DM component today, i.e. after freeze-out.


In the following we analyze the behavior of the relic abundance obtained by solving Eq.~\eqref{boltzc} making use of \texttt{micrOMEGAs 5.2.7a} \cite{Belanger:2014vza}, which calculates automatically all $\braket{\sigma v}$ in Eq.~\ref{boltzc}. We pay special attention to those regions in which $\Omega_s h^2$ acquires smaller values than in the SHP \cite{Cline:2013gha} for a fixed $(m_s,\lambda_{hs})$. This will be a necessary requirement in order to evade stringent direct detection bounds on the pseudoscalar DM from a few GeV to the TeV scale. Due to the subtle behavior of semi-annihilations in the $t$-channel, we divide the analysis in two parts, depending on the relative hierarchy between $m_s$ and $m_h$. Finally, for convenience, we introduce a new parameter $\Delta \equiv 2m_\psi - m_s - m_h$, such that as $\Delta \geq 0$ $s$-channel semi-annihilations are present in the relic abundance calculation. This is represented in blue in Fig.~\ref{parameter}~.

\begin{figure}[t!]
\centering
\begin{multicols}{3}
\includegraphics[width=0.3\textwidth]{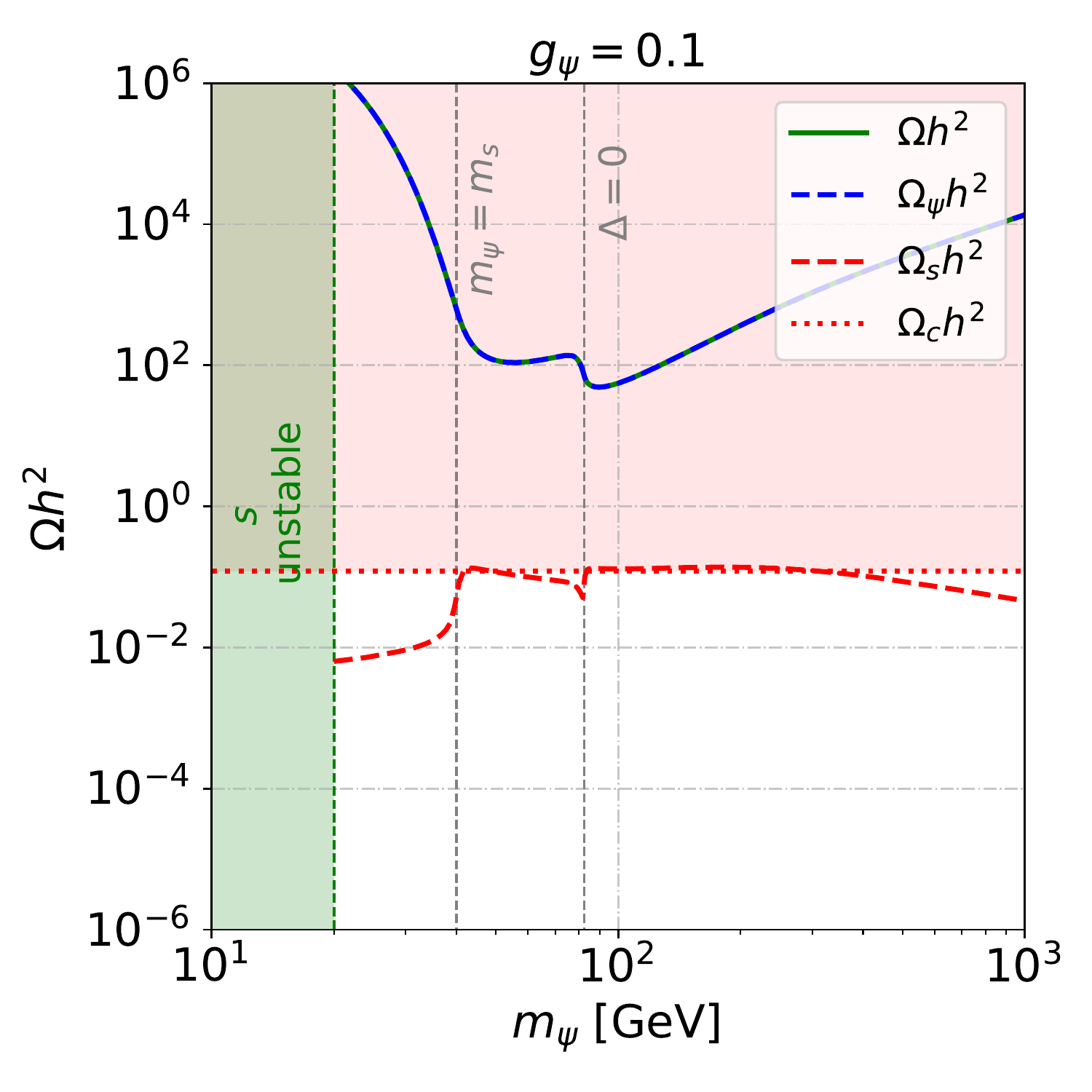}\par
\includegraphics[width=0.3\textwidth]{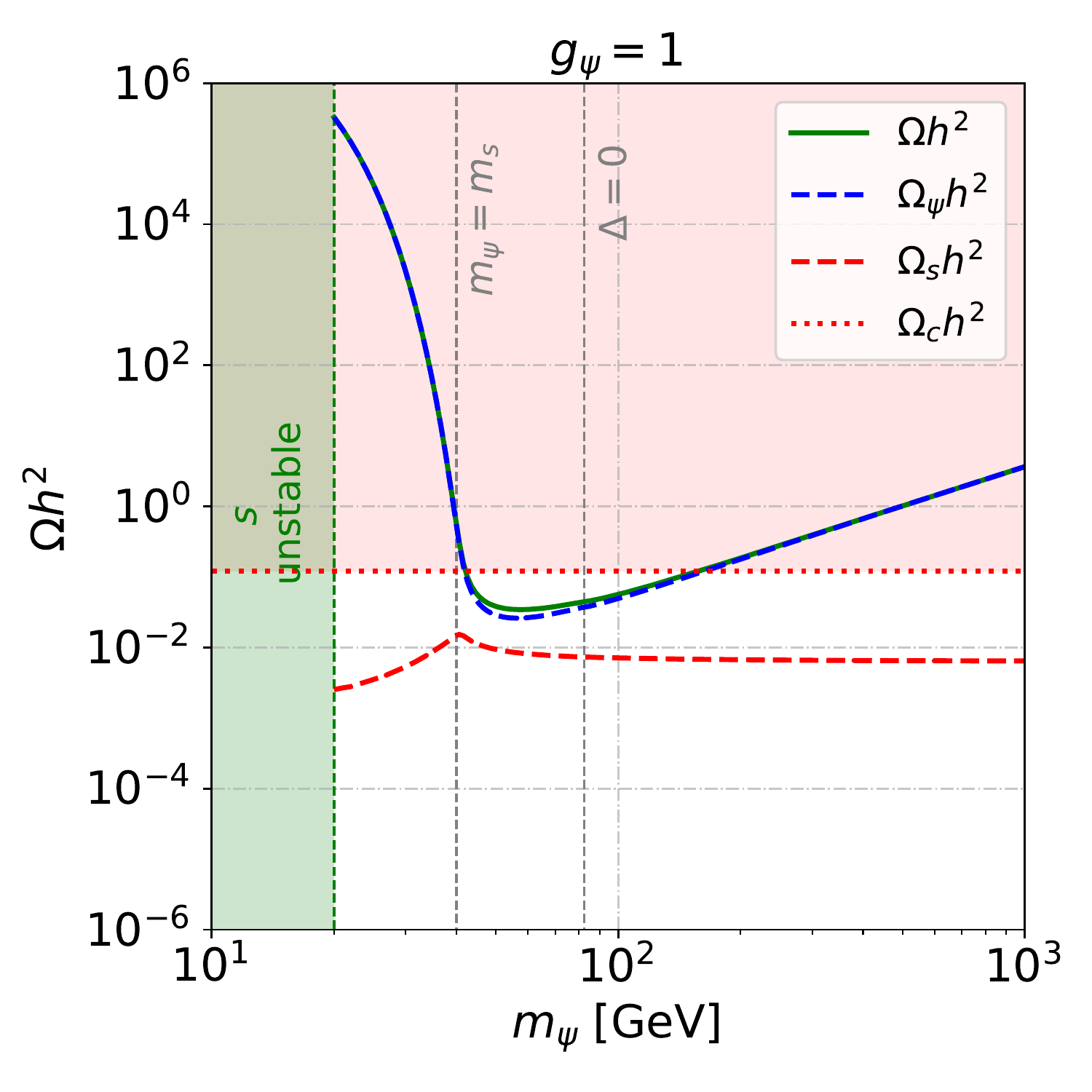}\par
\includegraphics[width=0.3\textwidth]{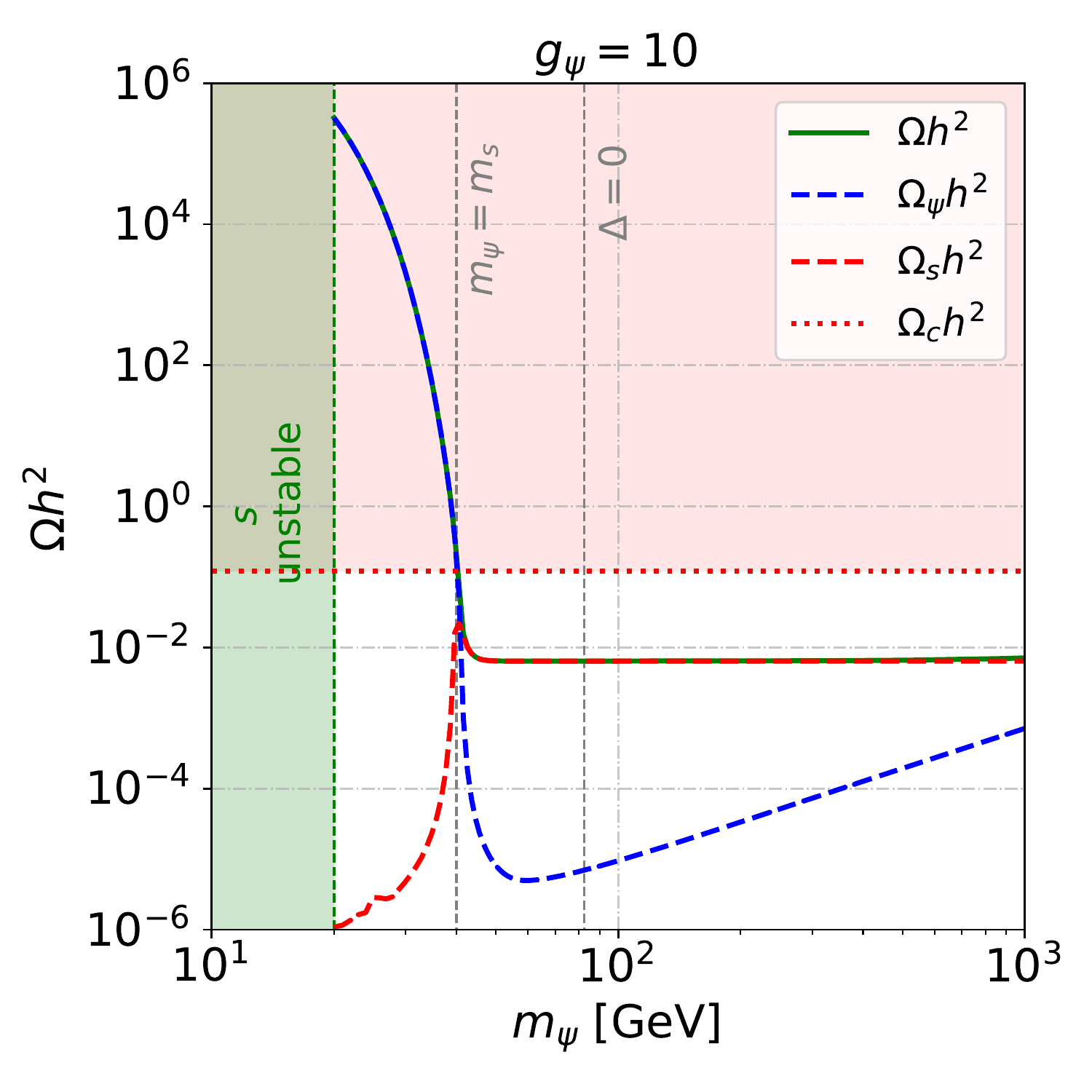}\par
\end{multicols}
\begin{multicols}{3}
\includegraphics[width=0.3\textwidth]{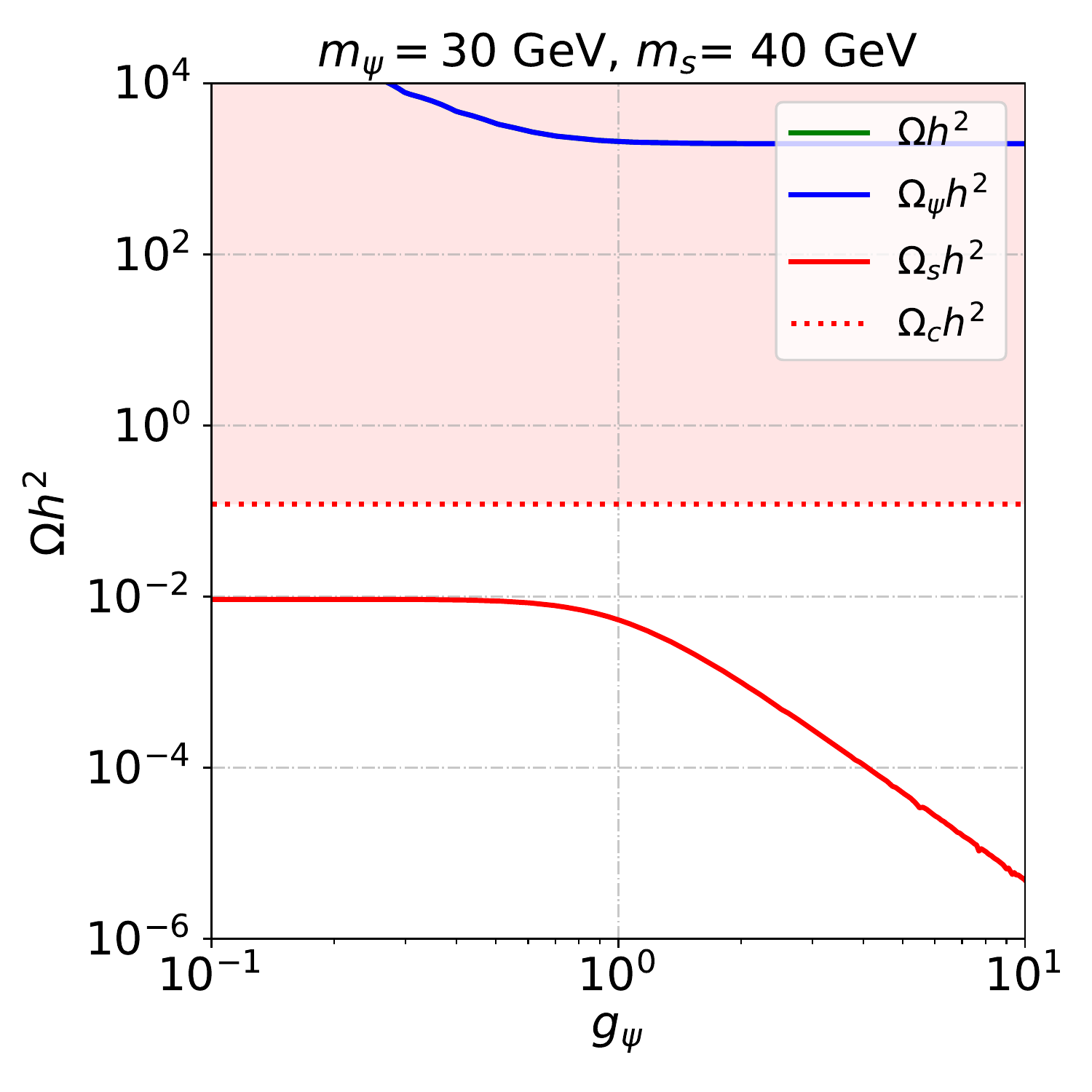}\par
\includegraphics[width=0.3\textwidth]{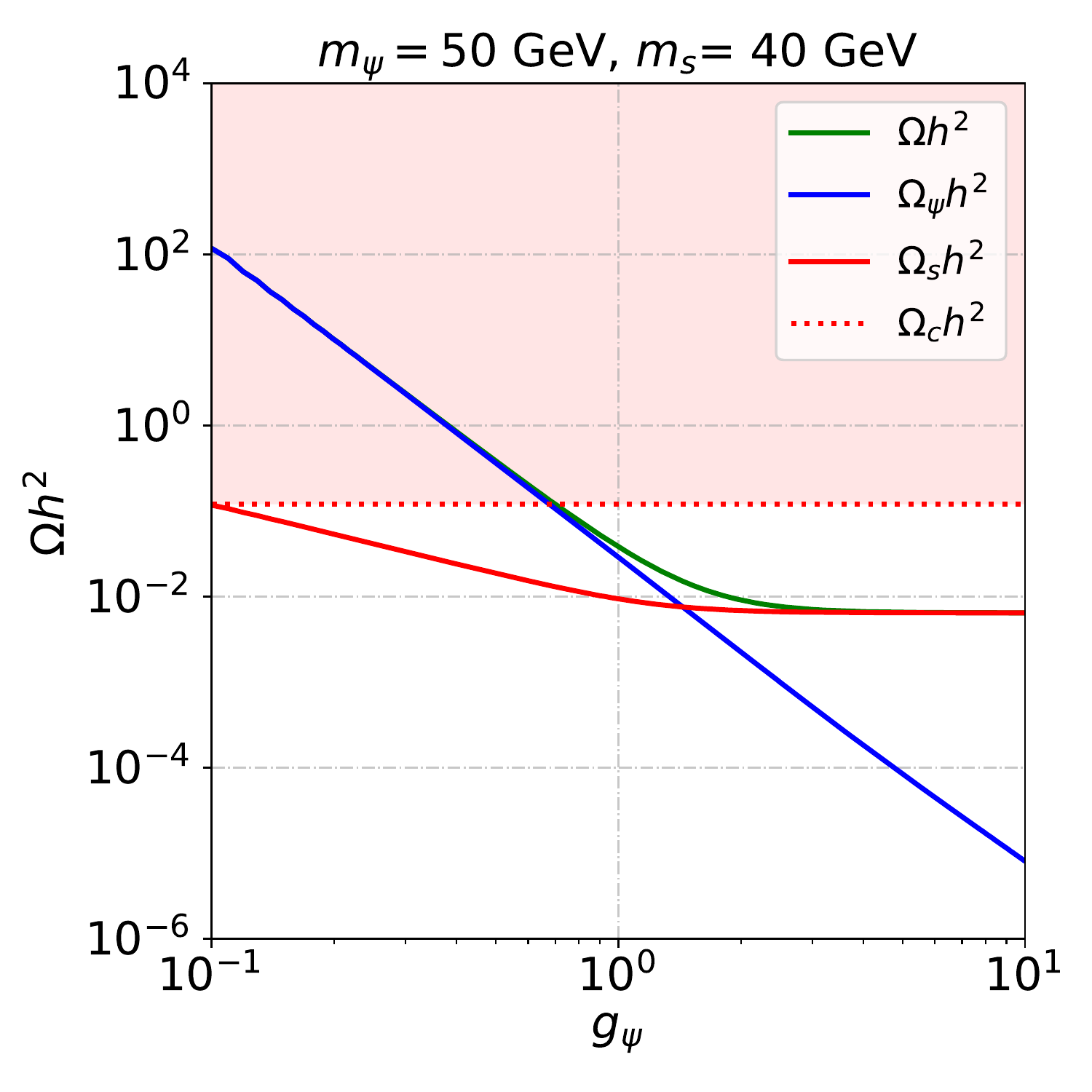}\par
\includegraphics[width=0.3\textwidth]{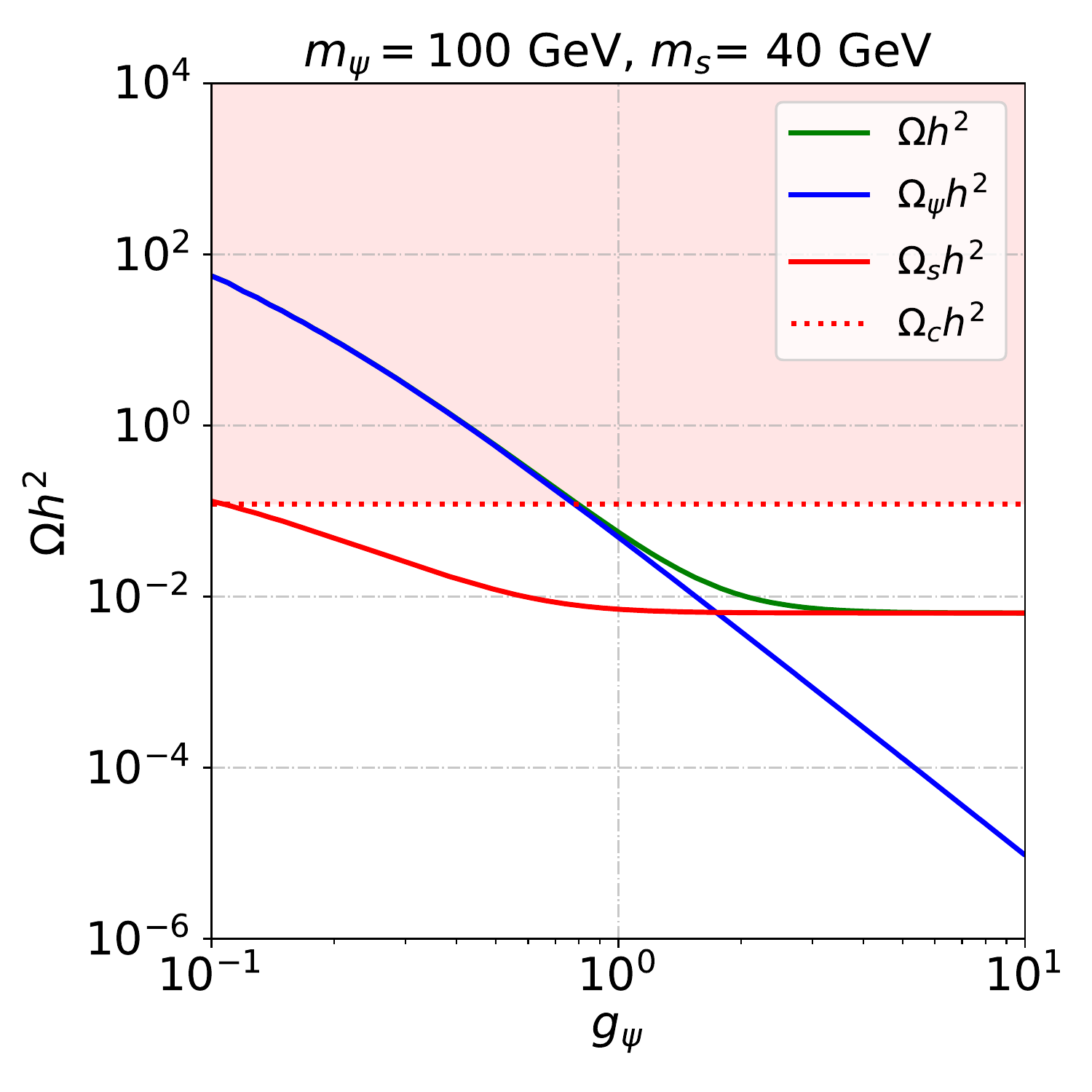}\par
\end{multicols}
\caption{\textit{(above) Relic abundance for $m_s = 40$ and $\lambda_{hs} = 1$, for different values of $g_\psi$ specified at the top of each plot. The green and red region are forbidden by the instability of $s$ and relic overabundance. The vertical dashed grey lines remark the points at which $m_\psi = m_s$ and $\Delta = 0$. \textit{(below)} Relic abundance as a function of $g_\psi$ for $m_s = 40$ and $m_\psi = 30, 40$ and $100$ GeV, respectively, with $\lambda_{hs} =1$.}}
\label{ev1}
\end{figure}
\subsubsection{$m_s < m_h$}\label{subsecint}
In this region all the processes in Fig.~\ref{diagrams} may be present but the semi-annihilation in the $t$-channel. Now, when $m_\psi$ is small enough, i.e.~$\Delta < 0$, then the $s$-channel semi-annihilation is not present either, and only pseudoscalar annihilations and DM conversions are present. Normally in this regime, an overabundance occurs as $m_s > m_\psi$, since $\psi$ does not have effective annihilation channels. As $m_\psi > m_s$, DM conversions of the type $\psi\bar{\psi}\rightarrow ss$ become effective, decreasing the overabundance of $\Omega_\psi$. If $m_\psi$ is big enough such that $\Delta \geq 0$, semi-annihilations in the $s$-channel become efficient, decreasing $\Omega_\psi$ even more. These characteristics can be seen in Fig.~\ref{ev1} (\textit{above}), where the abundances are depicted as a function of $m_\psi$, for $m_s = 40$ GeV, $g_\psi = (0.1, 1 ,10)$ and $\lambda_{hs} = 1$, with the green and red regions representing $s$ unstable and DM overabundance, respectively. The effects of conversions at $m_\psi = m_s$ becomes sharper as $g_\psi$ increases, since $\braket{\sigma_{\psi\psi ss} v}\sim g_\psi^4$ (see Appx.~\ref{app}). Equivalently, $s$-channel semi-annihilations near $\Delta = 0$ pushes down $\Omega_\psi$, and since both conversions and s-channel semi-annihilations depends inversely on $m_\psi^2$, their effectiveness decreases with the growing of $m_\psi$, then making $\Omega_\psi$ to increase for higher values of $m_\psi$. The changes of $\Omega_s$ are stronger only for $m_s > m_\psi$, where $ss\rightarrow \psi\bar{\psi}$ conversions are effective, and due to $\braket{\sigma_{ss\psi\psi} v} \propto g_\psi^4$, $\Omega_s$ decreases notoriously for such high couplings $g_\psi$ shown in the Fig.~\ref{ev1} (\textit{above}) right plot. We do not show the dependence of the relic abundances on $\lambda_{hs}$, but the changes are minor due to the fact that this parameter affects only the semi-annihilation intensity and modulates $\Omega_s$. 

In Fig.~\ref{ev1}~(\textit{below}) we show the abundances as a function of $g_\psi$, keeping $\lambda_{hs} = 1$. The relic abundance behavior depends strongly on the mass hierarchy and the magnitude of $g_\psi$. As it was previously discussed, for $m_\psi < m_s$ (left plot) it is $\Omega_\psi$ that dominates the relic, independently of $g_\psi$. In this case, $\lambda_{hs}$ would only change the relative abundance of the scalar singlet, but keeping very high values for $\Omega_\psi$. Oppositely, in the cases $m_\psi > m_s$ (middle and right plot), the relic hierarchy do depends on $g_\psi$, showing the effectiveness of conversions and $s$-channel semi-annihilation, respectively, with a notorious fall of $\Omega_\psi$ in each case.

\subsubsection{$m_s > m_h$}\label{subsecint2}
In this case, the $t$-channel semi-annihilation $\psi + s\rightarrow \psi + h$ is present, and it may participates strongly in the determination of $Y_s$ for $\Delta < 0$. The effectiveness of this semi-annihilation on $Y_s$ becomes highly sharp, due to the fact that once both DM components decouple from the SM plasma, $Y_s$ follows in a good approximation
\begin{eqnarray}\label{boltzd}
 \frac{dY_s}{dx} \propto - \frac{1}{2} x^{-2} \lambda_{s\psi\psi h}Y_\psi Y_s,\quad\quad x\gtrsim 10,             
\end{eqnarray}
\begin{figure}[t!]
\centering
\includegraphics[width=0.34\textwidth]{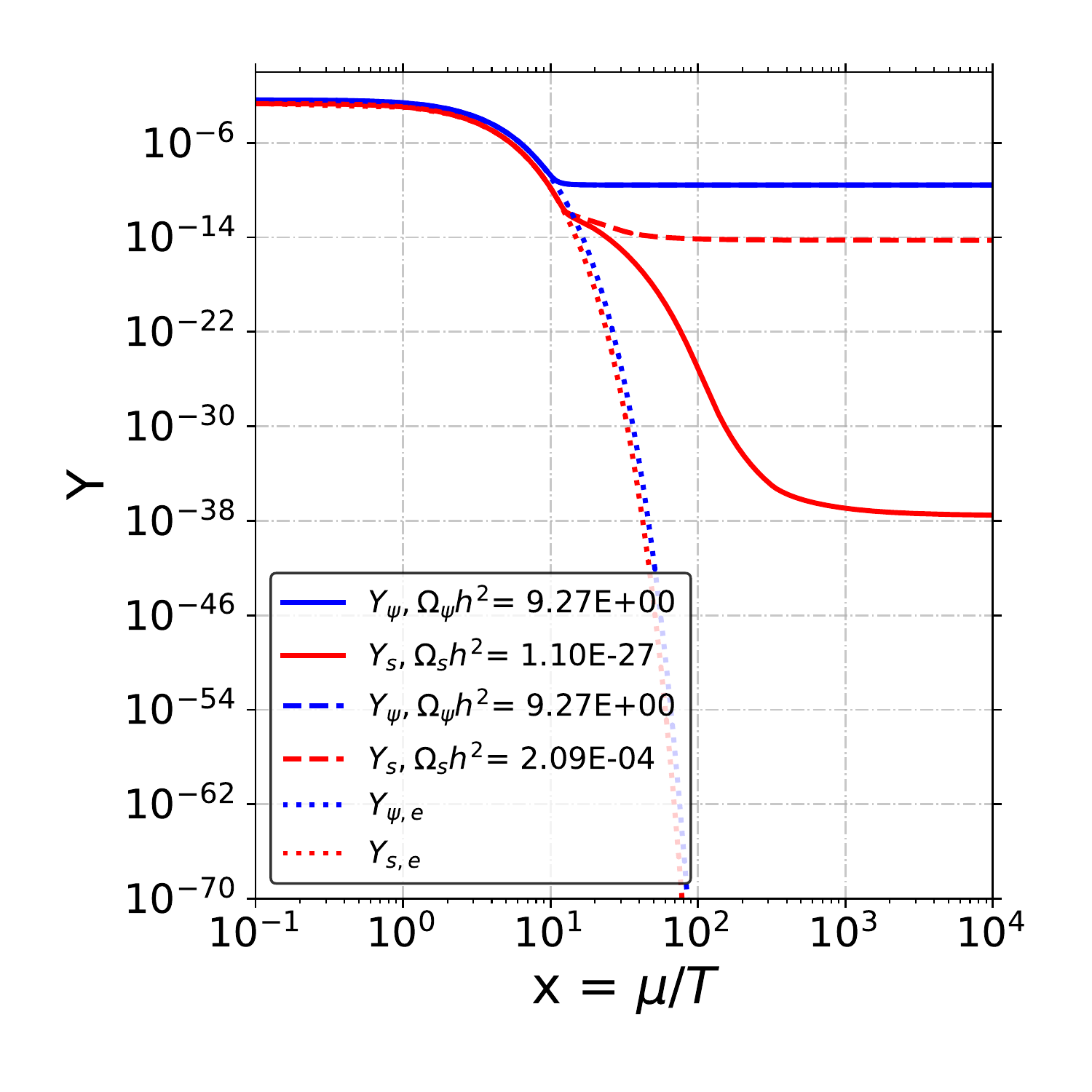}\quad
\includegraphics[width=0.34\textwidth]{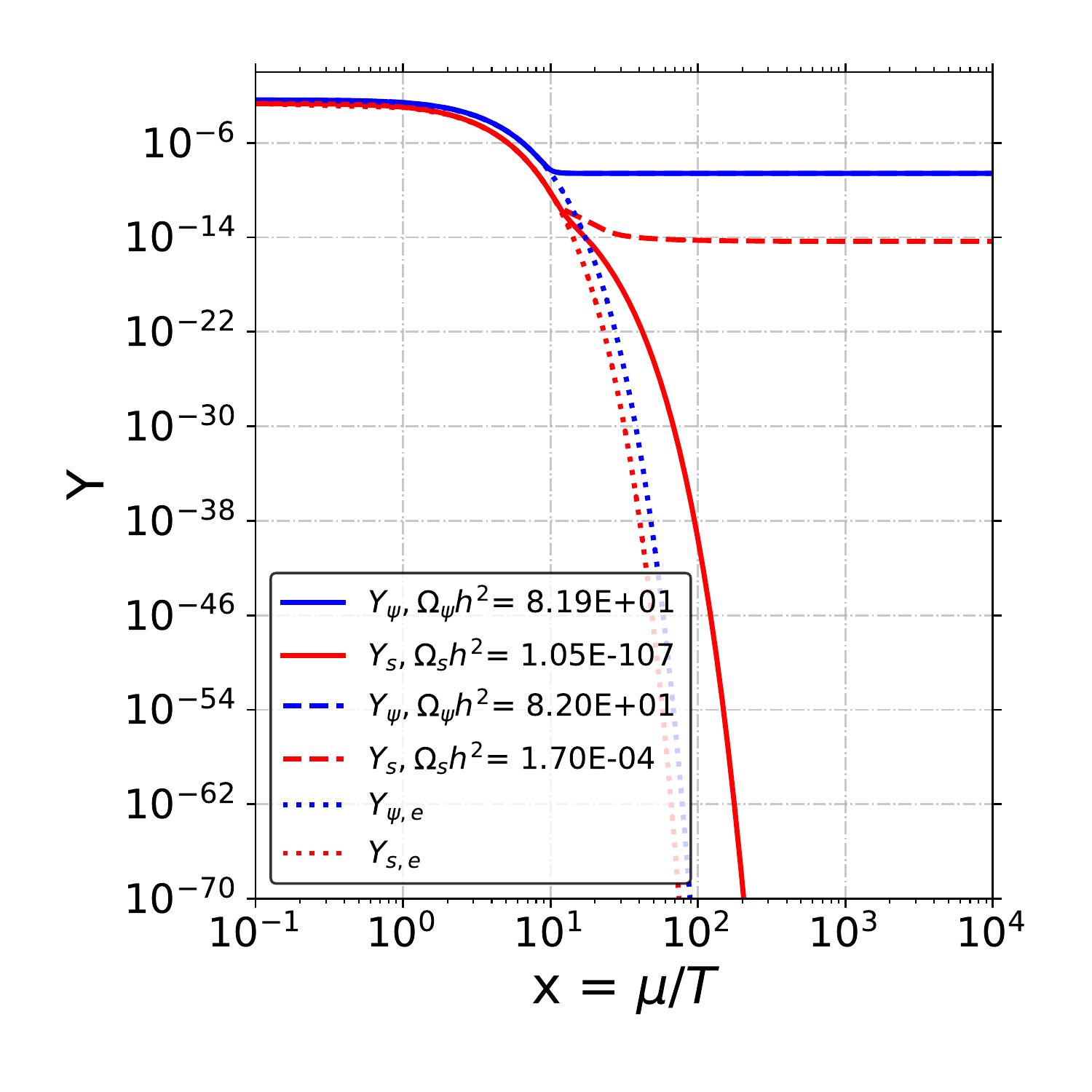}
\caption{\textit{DM yields as a function of $x$ in the region $m_s > m_h$ and $\Delta < 0$, considering (continuous lines) and not (dashed lines) the $t$-channel semi-annihilation. In both cases we have set $m_s = 130$ GeV and $g = \lambda_{hs} = 1$, with $m_\psi = 120$ GeV \textit{(left)} and $m_\psi = 110$ GeV \textit{(right)}. In the legends are specified the parameter densities for each case. The dotted lines correspond to the equilibrium densities of each singlet.}}
\label{semidiv}
\end{figure}
assuming $Y_\psi \approx$ constant and $Y_{s,e}\approx 0$. The solution of Eq.~\eqref{boltzd} gives an exponential suppression for $Y_s$, highly sensitive to the mass difference between the singlets. As an example of this behavior, in Fig.~\ref{semidiv} we show the evolution of the densities $Y_\psi$ (blue lines) and $Y_s$ (red curves) as a function of $x = \mu/T$, for $m_s = 130$ GeV, and $m_\psi = 120$ GeV (left plot) and 110 GeV (right plot). As Fig.~\ref{semidiv}\textit{(left)} suggests, $Y_s$ depends strongly on the $t$-channel semi-annihilation, as it can be compared the dashed and solid red curves, with the former not considering the process in the Boltzmann equation and the latter containing it. As the mass difference between the two singlets increases, the effects becomes sharper, as shown in Fig.~\ref{semidiv}\textit{(right)}. $Y_\psi$ is almost independent on this process, as it can be seen through the overlapping of the dashed and solid blue lines. This effect is particularly interesting due to the fact that as $\Omega_s$ becomes negligible, no direct detection bounds will apply on the pseudoscalar DM. This behavior was also seen in a two-component DM model consisting of two complex scalars stabilized by a $Z_5$ symmetry \cite{Belanger:2020hyh}. 

\begin{figure}[t!]
\centering
\includegraphics[width=0.34\textwidth]{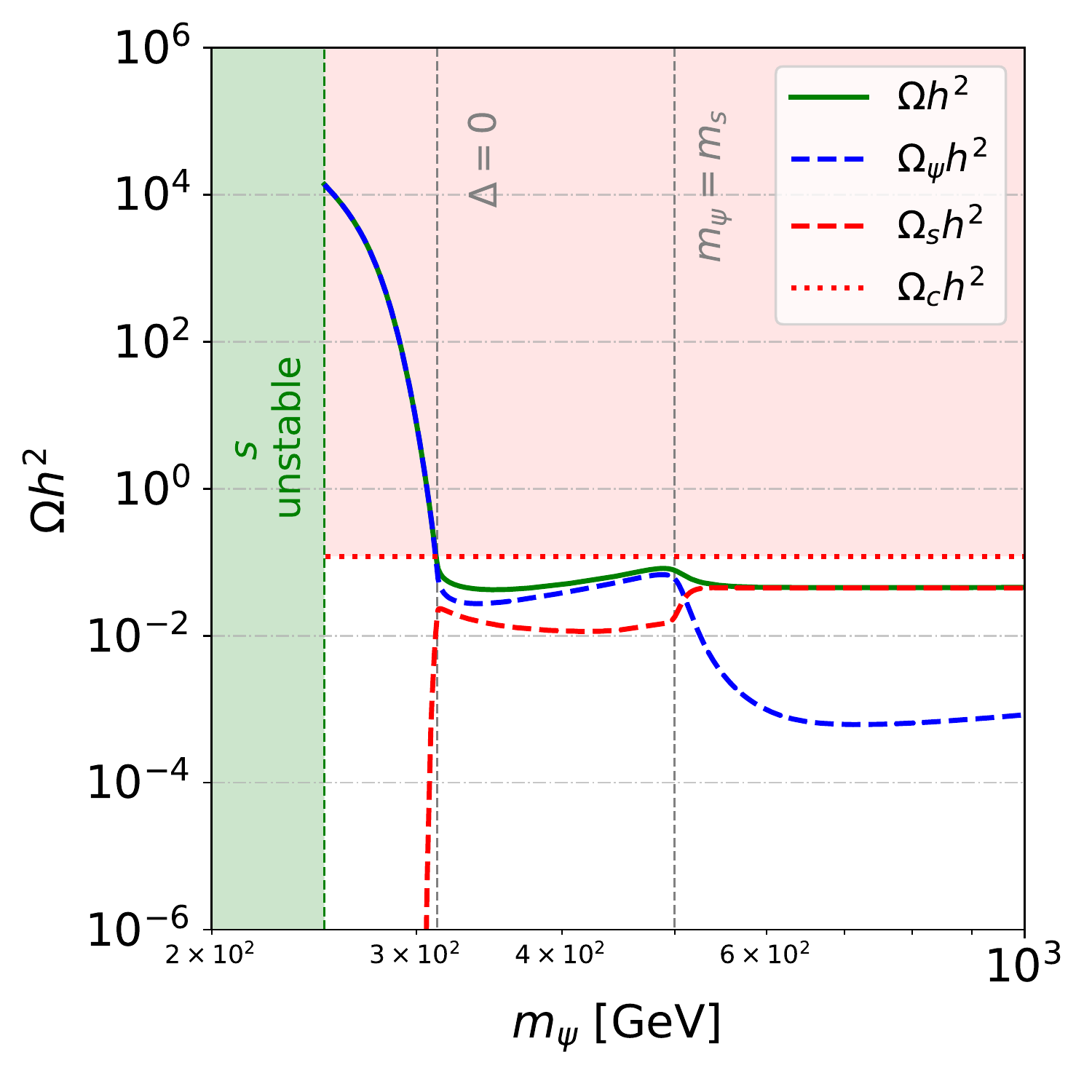}\quad\quad
\includegraphics[width=0.34\textwidth]{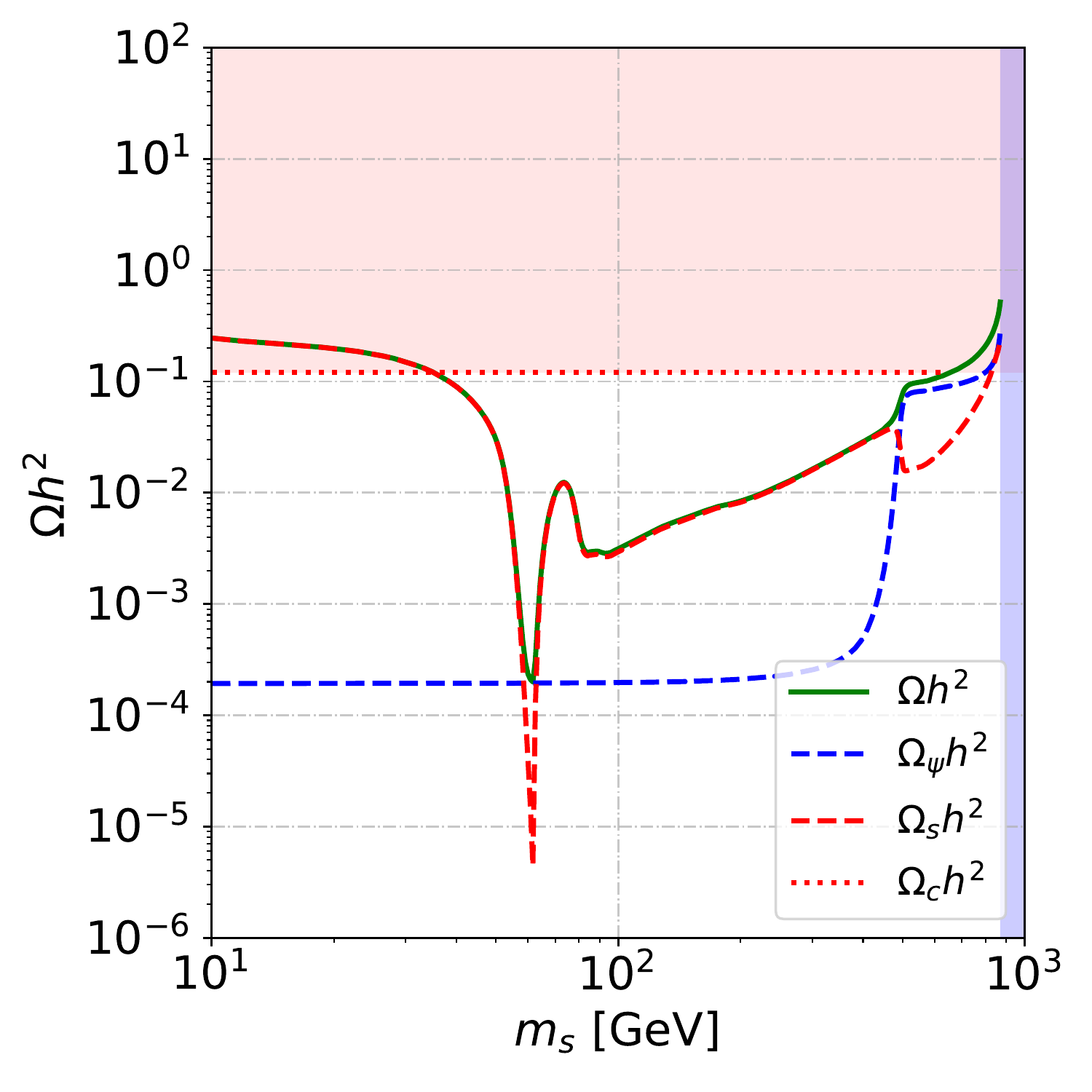}
\caption{\textit{(left) Relic abundance as a function of $m_\psi$, for $m_s =$ 500 GeV, $g_\psi = 10$ and $\lambda_{hs} = 0.25$. \textit{(right)} Relic abundance as a function of $m_s$, for $m_\psi = 500$ GeV, and the same couplings than in the left plot. The blue region in the right corresponds to $\Delta < 0$.}}
\label{ev2}
\end{figure}

For $\Delta > 0$, the two semi-annihilations enter into the coupled Boltzmann equations, and it is not longer possible to assume $Y_\psi$ constant, therefore Eq.~\eqref{boltzd} is not a good approximation to determine $Y_\psi$. In Fig.~\ref{ev2} we show the dependence of the relic abundance as a function of $m_\psi$ (left) and $m_s$ (right). Contrary to the case in the low mass regime of Fig.~\ref{ev1}, in the left plot of Fig.~\ref{ev2} we observe that conversions occur at higher values of $m_\psi$ than $s$-channel semi-annihilations. When the latter opens up, the total relic abundance decreases in various order of magnitude. There is an interesting effect we want to point out, namely the fall of $\Omega_s$ as $m_s \gtrsim m_\psi$. In both plots of Fig.~\ref{ev2} is possible to observe this effect, but in the right plot is more clear the role that conversions are playing, with the red dashed line making a tiny well from the soft growing as $m_s$ increases. This is, conversions of the type $ss\rightarrow \psi\psi$ start to be effective making a deviation from the well known behavior of the SHP \cite{Cline:2013gha} (for works with related behaviors see \cite{Belanger:2011ww, Maity:2019hre, Saez:2021oxl}). This type of wells, along with the Higgs resonance at $m_s \approx m_h/2$, will be of particular importance to evade direct detection constraints.  

In conclusion, we have analyzed the effectiveness of annihilations, conversions and semi-annihilations on the relic abundance of $\psi$ and $s$ in the parameter space. A random scan under theoretical and experimental constraints becomes useful in order to explore in major detail the viable regions of the model, and this is what we analyze in Sec.~\ref{phenosec}.


\section{Phenomenology}\label{phenosec}
In this section we present the relevant experimental constraints on the model along with a full scan of the parameter space. Once obtained the allowed parameter space, we explore indirect signals and we set upper limits from it. 
\subsection{Experimental Constraints}
When $m_s \leq m_h/2$, the Higgs boson can decay into two $s$ DM particles, with a decay width given by 
\begin{equation}\label{h11}
  \Gamma_{inv}(h\to ss) = \frac{\lambda^2_{hs}v_H^2}{32\pi m_h}\sqrt{1 - \frac{4m^2_s}{m^2_h}}.
\end{equation}
This contributes to the invisible branching rate $\text{Br}(h\to\text{inv})=\Gamma_\text{inv}/(\Gamma_\text{SM}+\Gamma_\text{inv})$, where the total decay width of the Higgs into SM particles is given by $\Gamma_\text{SM}~=~ 4.07$~ MeV\cite{Sirunyan:2018owy}. Experimental searches put stringent constraints on this quantity, and the most strict value is given by $\text{Br}(h\to\text{inv}) < 0.19$ at $95\%$ C.L. \cite{Sirunyan:2018owy}.

From the DM sector we have the following constraints. First, the measurement of the DM relic density today, given by $\Omega_c h^2 = 0.120\pm 0.001$ \cite{planck2018}. Based on the uncertainties in our computation, we apply this constraint with a tolerance of $\sim~$10\%, i.e. $\Omega_c h^2 \in [0.11, 0.12]$. Secondly, direct detection, which in multicomponent DM scenarios the interaction of DM with nucleus matter through the spin-independent (SI) cross section comes from weighting it by a factor which takes into account the relative abundance $\Omega_i$ over the Planck measured $\Omega_c$, i.e.,
 \begin{eqnarray}\label{sigmahat}
  \hat{\sigma}_\text{SI,i} = \left(\frac{\Omega_i}{\Omega_{\text{c}}}\right)\sigma_\text{SI,i},\quad i=\psi,s,
 \end{eqnarray}
where $\sigma_\text{SI,i}$ is the spin-independent cross section. At tree level, it is only the pseudoscalar DM which interacts with nuclei via the Higgs portal, with its SI cross section given by \cite{Cline:2013gha}
\begin{equation}
  \sigma_\text{SI,s}=\frac{\lambda^2_{hs}f^2_N}{4\pi}\frac{\mu_i^2m^2_n}{m^4_hm_s^2},
\end{equation}
with $m_n$ as the nucleon mass, $f_N$ is a factor proportional to the nucleon matrix elements, and $\mu~=~m_nm_s/(m_n+m_s)$ is the DM-nucleon reduced mass. Upper bounds on $ \sigma_\text{SI}$ are given by XENON1T data \cite{Aprile_2018} and the projections of XENONnT \cite{Aprile:2015uzo}. 

\subsection{Scan}
\begin{figure}[t!]
\centering
\includegraphics[width=0.43\textwidth]{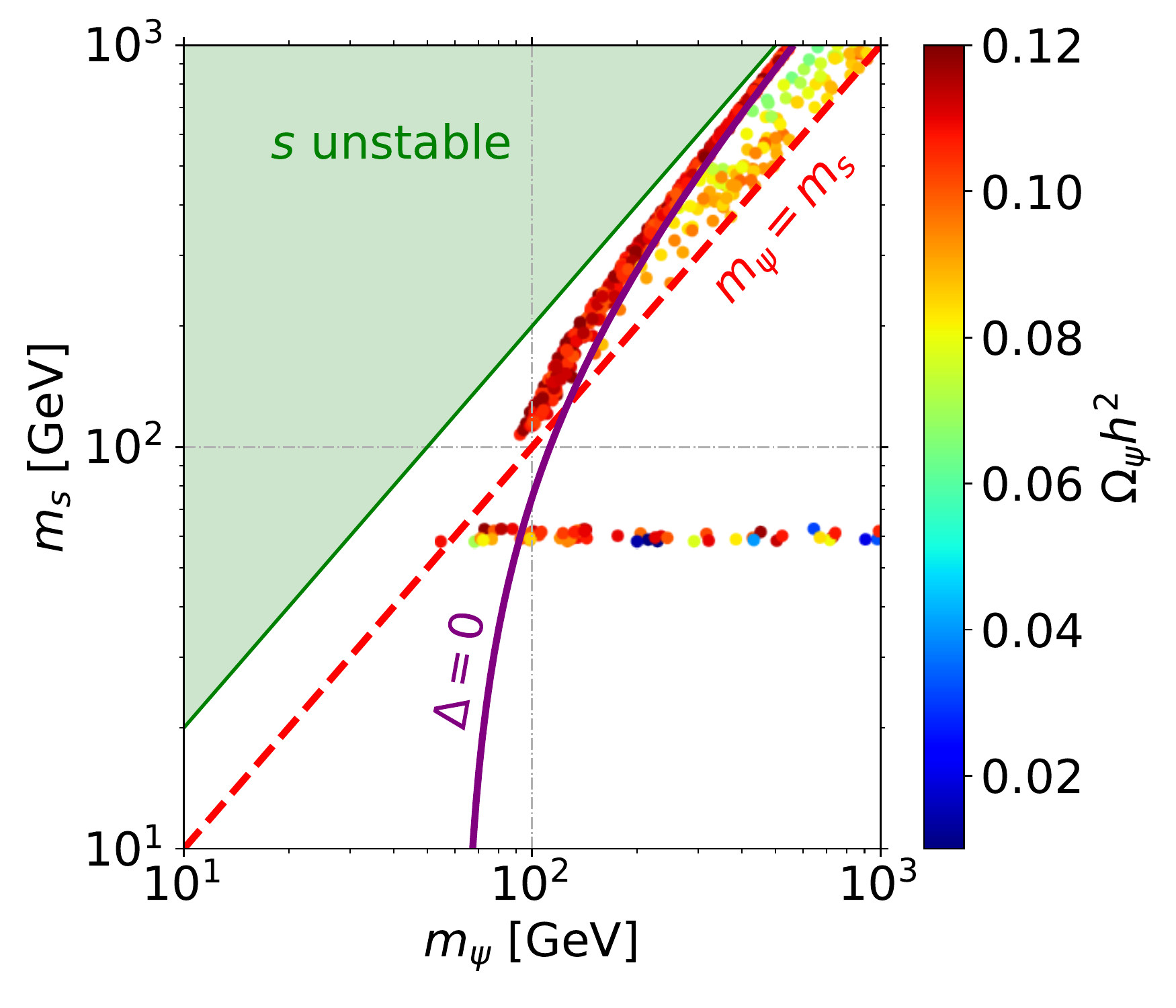}\quad
\includegraphics[width=0.43\textwidth]{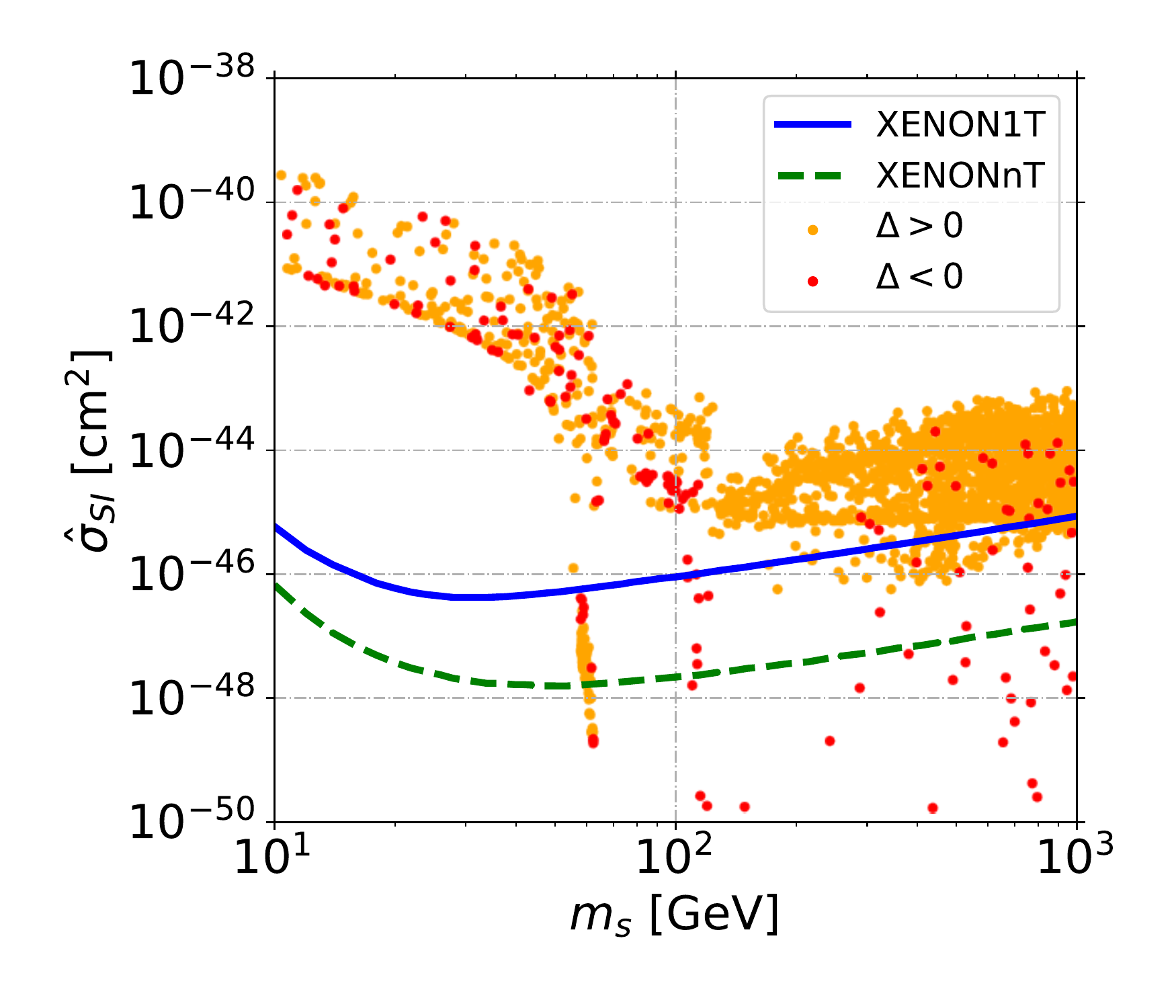}
\caption{\textit{(left) Points fulfilling direct detection and relic abundance in all the scanned parameter space. \textit{(right)} The same selected random points shown in the left plot, now projected on $(m_s , \hat{\sigma}_{SI,s})$, where we distinguish points with $\Delta > 0$ (orange) and $\Delta < 0$ (red). The continuous blue line corresponds to the upper limit given by XENON1T (1 t$\cdot$y)\cite{Aprile_2018}, whereas the green dashed line a projection for XENONnT (20 t$\cdot$y) \cite{Aprile:2015uzo}.}}
\label{sighatm2}
\end{figure}

In this section we show a scan on the parameter space in the range $m_{\psi,s} \in [10,1000]$ GeV (subject to $m_s < 2m_\psi$) and  $g_\psi,\lambda_{hs} \in [0.001,4\pi]$. The full scan highlight interesting features that were already anticipated in Sec.~\ref{2dmsec}. In Fig.~\ref{randomscans1} we show the scan of points projected on the plane $(m_\psi , m_s)$, with the density color representing the total abundance $\Omega h^2$ (left), $\Omega_\psi h^2$ (middle) and $\Omega_s h^2$, respectively. The green region (top left) corresponds to those points with $m_s > 2m_\psi$, then making $s$ unstable. We pay attention to four regions based on Fig.~\ref{randomscans1} (\textit{left}): 
\begin{figure}[t!]
\centering
\includegraphics[width=0.3\textwidth]{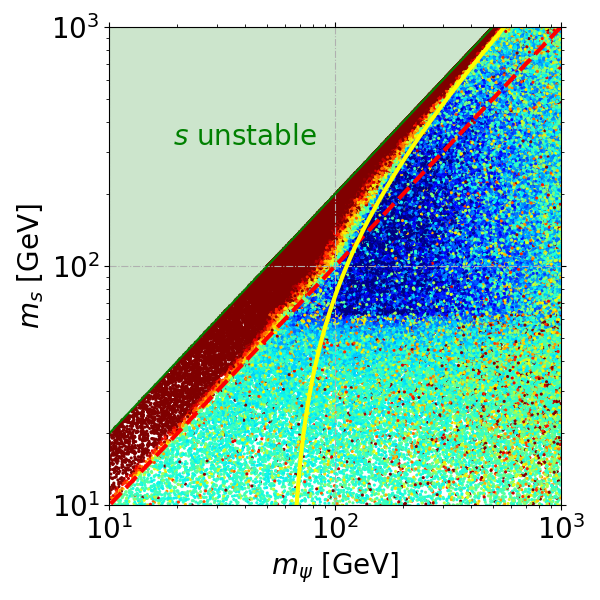}
\includegraphics[width=0.3\textwidth]{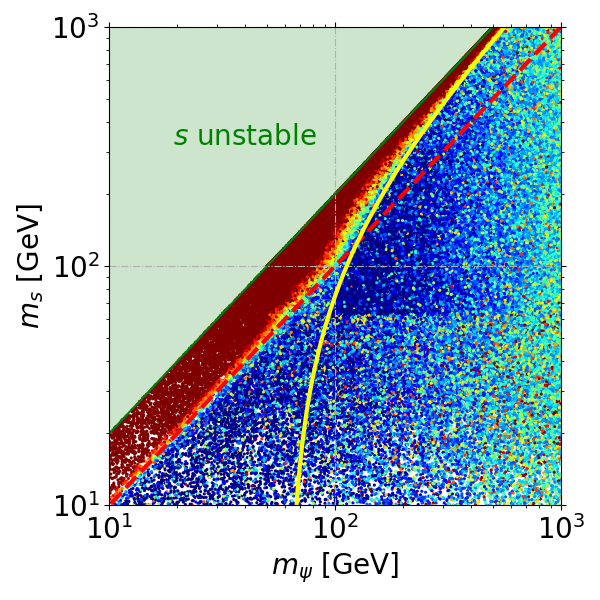}
\includegraphics[width=0.35\textwidth]{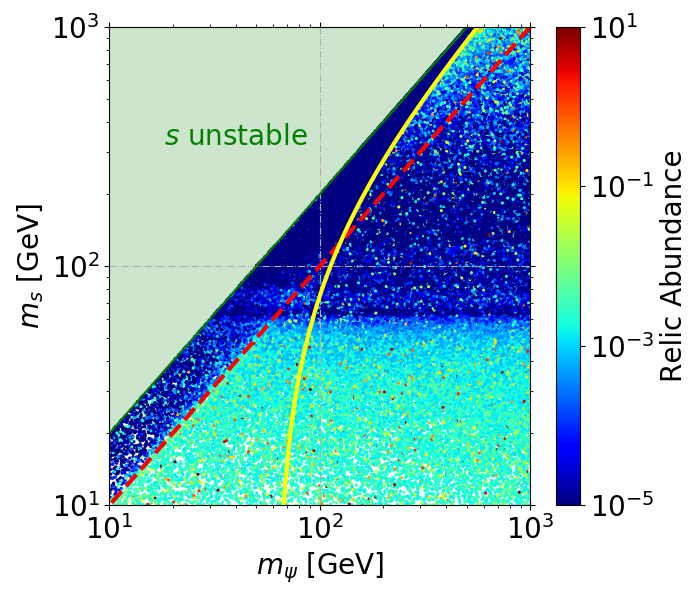}
\caption{\textit{Random scan points projected in the mass plane with the bar color indicating the total relic abundance $\Omega h^2$ \textit{(left)}, $\Omega_\psi h^2$ \textit{(middle)} and $\Omega_s h^2$ \textit{(right)}. The homogeneous green region in the top left of each plot corresponds to $s$ unstable, i.e. $m_s > 2m_\psi$. The red dashed line indicates $m_\psi = m_s$ and the yellow one the frontier $\Delta = 0$.}}
\label{randomscans1}
\end{figure}
\begin{enumerate}[label=(\alph*)]
 \item \textit{Dark red band}. In general, $\Omega_\psi$ get too big values for $m_\psi < m_s$, since no effective annihilation channels for the singlet fermion are present (see Sec.~\ref{subsecint}). However, there is a smooth relic density transition from the deep red to lower relic densities, due to the thermal tail distribution of conversions and $s$-channel semi-annihilation as $\Delta \lesssim 0$. With respect to $\Omega_s$, it is always sub-abundant, specially for $m_s > m_h$ where the $t$-channel semi-annihilations are present, and as we have described in Sec.~\ref{subsecint2}, tiny mass shift between $\psi$ and $s$ makes $\Omega_s$ to decrease strongly (deep blue region in Fig.~\ref{randomscans1} (\textit{right})).   
 \item \textit{Green area in below}. In this region the relic is mostly dominated by the pseudoscalar, as it can be seen from the plots. As it was shown in Fig.~\ref{ev1}, $\Omega_\psi < \Omega_s$ for conversion-driven processes of the type $\psi\bar{\psi}\rightarrow ss$ and moderate couplings ($g_\psi \sim 1$). For very high $m_\psi$ is possible to observe that $\Omega_\psi$ tend to increase, as conversions and semi-annihilations are less effective. 
 \item \textit{Higgs resonance and thresholds}. In the region 50 GeV $\lesssim m_s \leq m_h$ the Higgs resonance and SM thresholds are present, then pseudoscalar annihilations are enhanced. These effects are more clear in Fig.~\ref{randomscans1} (\textit{right}) for $\Omega_s$. For $m_s~\lesssim~m_h/2$, $s$-channel annihilations of the singlet pseudoscalar via Higgs resonance tend to be very effective, decreasing $\Omega_s$ substantially, as it can be seen with the deep blue horizontal line. For $m_h/2 < m_s < m_W$ the abundance of $s$ lift up, and for $m_s \lesssim m_W$ the relic decreases again. The thresholds at $m_Z$ and $m_h$ are also present but they are less notorious. 
 \item \textit{Blue-green upper region}: For the singlet masses $\gtrsim$ 100 GeV, the relic of both components tends to be low (blue points), but as the masses increase the effectiveness of conversions and semi-annihilations become less strong, i.e. $\braket{\sigma v} \sim m_{\psi,s}^{-2}$ (Appx.~\ref{app}), then raising the relic of both components. In this mass region the relic abundance seems to acquire moderate values in order to fulfill the Planck measurement.
\end{enumerate}
The conjunction of the three plots suggests that most of the points that could give the correct relic abundance are just in very specific sectors, and those regions must to shrink even more when other constraints are taken into account. In Fig.~\ref{sighatm2}~\textit{(left)} we show the resulting parameter space points after imposing the constraints from relic abundance, Higgs to invisible and direct detection bounds. Specifically, the regions are: (i) Higgs resonance, (ii) region with $m_s > m_h$ and $\Delta < 0$, and (iii) in the high mass with $\Delta > 0$ and $m_\psi < m_s$, indicating the corresponding $\Omega_\psi$ with a color bar. Region (i) is the only one that allows $m_\psi > m_s$ and where the Higgs to invisible constraint applies \footnote{In the presence of a resonance, the smallness of the Higgs portal coupling may breakdown the assumption of local thermal equilibrium when chemical decoupling is taking place. In our case we have not considered those effects, where a more involved treatment must be carried \cite{Binder:2017rgn}.}. The points in region (ii) easily evade XENON1T constraints due to the fact that $\Omega_s\rightarrow 0$, then the condition $\Omega_\psi = \Omega_c$ is enough to fulfill all the experimental constraints. Finally, in the high mass region we found some points anticipated by the analysis around Fig.~\ref{ev2}, where the power of conversions and $s$-channel semi-annihilations makes $\Omega_s$ to decrease enough evading XENON1T. The cost of the latter implies values for $g_\psi$ near the perturbativity limit criteria (this point is discussed in the last section). In Fig.~\ref{sighatm2}~(right) we project the scan points on the plane $(m_s , \hat{\sigma}_{SI,s})$, contrasted with the limits given by XENON1T and the XENONnT projection \cite{Aprile:2015uzo}. Even when XENON experiments rule out most of the orange points (i.e. regions (i) and (iii), a fraction of red points ($\Delta < 0$ and $m_s > m_h$) easily evade the strongest direct detection constraints. It is worth to say that this inverted peak at the Higgs resonance is a typical characteristic of Higgs portals, hardly to be ruled out by experiments. 

To end this subsection, we would like to point out about the limitations of our random scan, which was based on overlaying exclusion limits (mainly from relic abundance and direct detection). It is well known that this type of exclusions present some drawbacks, such as the blindness to the influence of the choice of some parameters (e.g. DM halo distribution or the Higgs mass pole) on the experimental limits, or the fact that the \textit{allowed regions} do not present additional information about which points are more favored than others. Considering those limitations, statistical analysis based on combined likelihoods functions may give a more accurate and realistic way to cure those disadvantages, although beyond the scope of the present analysis \cite{GAMBIT:2017gge, AbdusSalam:2020rdj}.

\subsection{Indirect detection}
In the previous subsection we found that three regions of the parameter space fulfill invisible Higgs decay, relic abundance and direct detection constraints, with two of them being practically in the region $\Delta > 0$. In that case, $s$-channel semi-annihilations are expected to give sizable fluxes of particles today through its $s$-wave nature. The $t$-channel semi-annihilation also goes in the $s$-wave, but we have checked that in general its cross section is lower than the corresponding $s$-channel, and indirect signals with a pair of $s$ in the initial state have shown to be not sizable outside the Higgs resonance \cite{Cline:2013gha}. Additionally, $\Omega_\psi h^2$ reaches $\sim 50\%$ to $99\%$ of the total DM budget in the high mass regime (see Fig.~\ref{sighatm2}(left)), consequently the scaling factor $(\Omega_\psi/\Omega_c)^2$ does not considerable suppress the flux produced by the $s$-channel semi-annihilation. In the following, we show the box-shaped differential spectra \cite{Ibarra:2012dw} for this channel with its subsequent decay $h\rightarrow \gamma\gamma$, and secondly, restrictions on the parameter space coming from bounds based on searches of gamma rays (Fermi-LAT), anti-protons (AMS-02), and projections from the Cherenkov Telescope Array (CTA) are presented.
\subsubsection{Box-shape gamma ray}
The differential flux of photons produced in fermionic DM annihilations and received at earth from a given solid angle in the sky $\Delta\Omega$ with a detector of area $A$ is given by
\begin{eqnarray}
 \frac{d\Phi_\gamma}{dE_\gamma} = \frac{1}{A}\frac{dN_\gamma}{dE_\gamma dt} = \frac{\braket{\sigma _{sh}v}}{8\pi m_{\psi}^2}\left(\text{Br}_{h} \frac{dN}{dE_\gamma}\right) \left(\frac{1}{\Delta\Omega}\int_{\Delta\Omega} Jd\Omega\right),
\end{eqnarray}
where $\braket{\sigma_{sh} v}$ is the corresponding average annihilation cross section times velocity of the process ${\psi + \bar{\psi}\rightarrow s + h}$, $\text{Br}_h$ is the branching ratio of the Higgs into two photons, and $dN/dE_\gamma$ the corresponding normalized spectra. The $J$ factor is the integral of the squared DM density $\rho_{DM}$ along the line of sight $J = \int_{\text{l.o.s}} ds\rho^2_{DM}$. We consider as our main region of interest the galactic center, which features $\Delta\Omega = 1.30$ sr, $\int_{\Delta\Omega} Jd\Omega = 9.2\times 10^{22}$ GeV$^2$cm$^{-5}$, assuming a NFW profile normalized to a local DM density of 0.4 GeV/cm$^3$ \cite{Vertongen:2011mu}. To determine the normalized spectra of emitted photons, we first note that their energies, in the fermion DM collision frame\footnote{Today, fermion DM $\psi$ moves non-relativistically, therefore colliding practically in the earth rest frame.}, are given by
\begin{eqnarray}\label{photons}
 E_{\gamma,1} = \frac{m_h^2/2}{E_h - \sqrt{E_h^2 - m_h^2}\cos\theta},\quad \quad E_{\gamma,2} = E_h - E_{\gamma,1},
\end{eqnarray}
where $\theta$ corresponds to the angle sustained by one of the photons and the in-flight Higgs. The energy of the emitted scalar particles in the rest frame of the collision of $\psi$ are given by
\begin{eqnarray}\label{energies}
 E_h = m_\psi\left(1 - \frac{m_s^2 - m_h^2}{4m_\psi^2}\right),\quad E_s = m_\psi\left(1 + \frac{m_s^2 - m_h^2}{4m_\psi^2}\right). 
\end{eqnarray}
For a fixed $E_h$, the energy of the photon received at earth depends only on $\theta$, with a maximum (minimum) energy for $\theta = 0 (\pi/2)$ \footnote{It is possible to receive the two photons at earth as $\theta\rightarrow 0$, or equivalently, when both photons are emitted transverse to the Higgs direction. This scenario requires a highly boosted Higgs and a detailed analysis of the emitted photons.} then displaying a box-shaped spectrum centered at $E_c \equiv (E(0) + E(\pi))/2 = E_h/2$ and with a width $\Delta E \equiv E(0) - E(\pi) = \sqrt{E_h^2 - m_h^2}$. Therefore, the normalized spectra can be written as
\footnote{Note that in the $\psi$ annihilation center-of-mass frame, due to the conservation of angular momentum, the two photons must be emitted back to back if they have the same polarization, and co-linearly if they have opposite polarization; the conservation of linear momentum requires $s$ to be emitted along with one of the photons in the former case, and in the direction opposite to the photons in the latter. See also \cite{Ghosh:2019jzu}.}. 
\begin{eqnarray}
 \frac{dN}{dE_\gamma} = \frac{2}{\Delta E}\Theta\left(E_\gamma - E_c + \frac{1}{2}\Delta E\right)\Theta\left(E_c +\frac{1}{2}\Delta E - E_\gamma\right),
\end{eqnarray}
where the factor multiplicative factor 2 accounts for the two emitted photons (then $E_\gamma$ any of the two photons in \ref{photons}), and $\Theta$ are Heaviside functions. 

\begin{figure}[t!]
\centering
\includegraphics[width=0.40\textwidth]{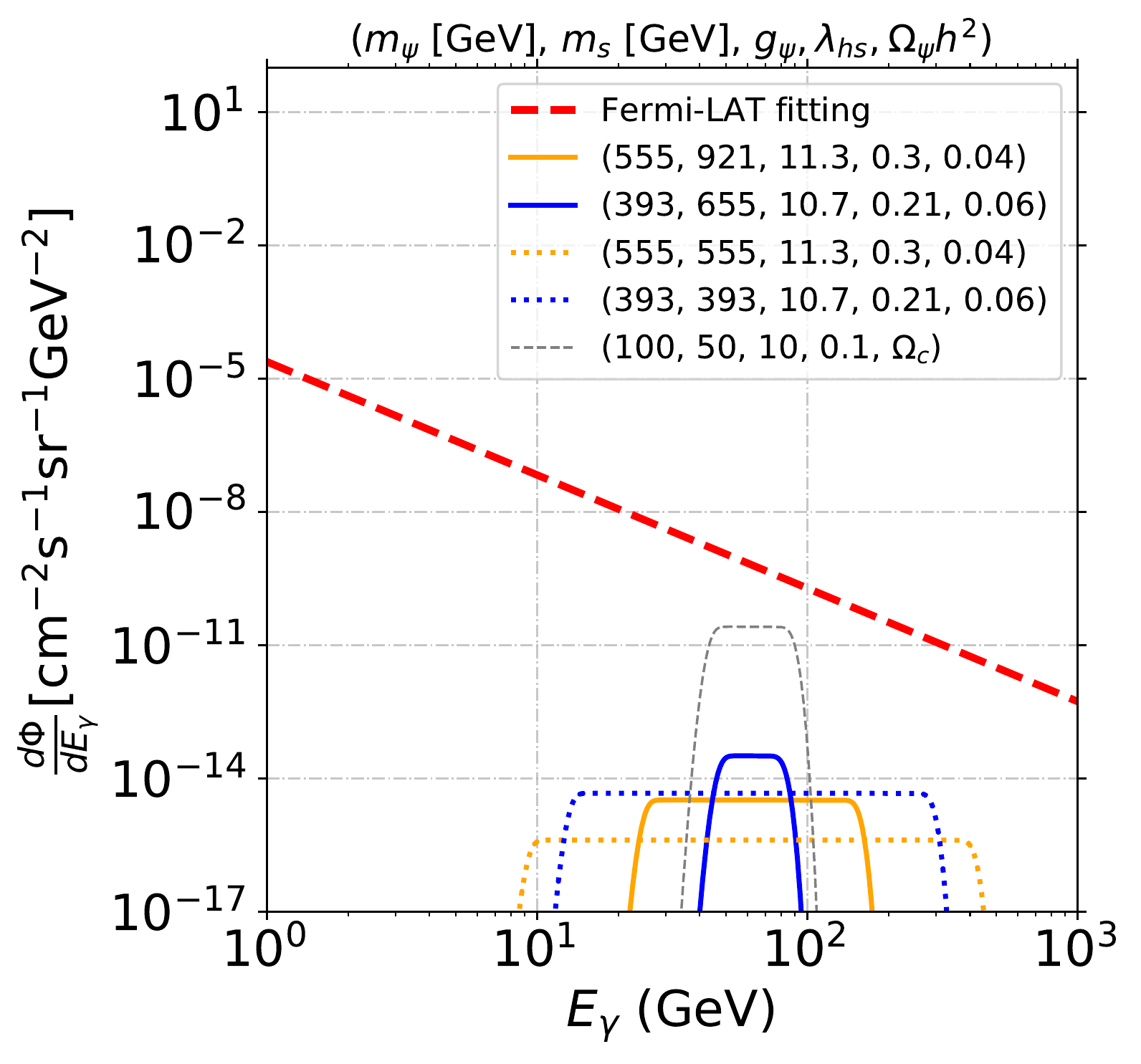}
\caption{\textit{Differential gamma ray flux originated from the process $\psi + \bar{\psi}\rightarrow s + h$, followed by the decay $h\rightarrow \gamma\gamma$, for two random points fulfilling relic abundance and direct detection (continuous lines), whereas the dotted curves represent hypothetical points in which the two DM components are completely degenerated. The dashed red line corresponds to the flux given by Fermi-LAT \cite{Ibarra:2012dw}, and the gray an hypothetical point with low $(m_\psi , m_s)$.}}
\label{flux_micro}
\end{figure}

In Fig.~\ref{flux_micro}~(\textit{left}) we show the box-shaped spectra for two random points allowed by the relic density abundance and XENON1T, contrasted with the data given by Fermi-LAT (red dashed line), where we have taken the background fitting function in order to estimate the data signal: $d\Phi_\gamma/dE = 2.4\times 10^{-5}(E_\gamma/$GeV$)^{-2.55}$ GeV$^{-1}$cm$^{-2}$s$^{-1}$sr$^{-1}$ \cite{Vertongen:2011mu}. As it was anticipated above, the differential fluxes for the selected points are well below the red dashed line, in principle not showing any possible tension. The blue and orange dot curves show the spectra when the two DM components become degenerated, widening the boxes and therefore peaking at higher energy photons. Only points with masses (and high couplings) as small as $m_\psi \sim 100$ GeV may approach significantly to the red dashed line, as it is depicted by the grey dashed line in Fig.~\ref{flux_micro}~(\textit{left}). It is known that the power of box-shaped gamma rays goes in their potential deviation with respect to the power-law Fermi-LAT background, then the previous result is premature to discard a possible constraint on the parameter space. In order to do this, in the following we use the recent upper bounds on gamma rays produced in semi-annihilations based on Fermi-LAT data, along with anti-protons flux measurements.

\subsubsection{Upper bounds}
Upper limits on $\braket{\sigma_{hs} v}$ have not been constructed yet in order to constraint the parameter space via indirect searches. However, they can be inferred from existing upper bounds on the average annihilation cross section times velocity for $DM+DM\rightarrow DM+h$ process based on gamma-ray signals \cite{Queiroz:2019acr}, and from $DM+DM\rightarrow b\bar{b}(W^+W^-)$ based on anti-proton flux measurements \cite{Reinert:2017aga}. In the following we derive the useful algebraic relations to translate the existing upper bounds to our average semi-annihilation cross section times velocity $\braket{\sigma_{sh} v}$.

First, let us consider the process $\psi\bar{\psi}\rightarrow \psi 'h$, with $\psi$ and $\psi'$ arbitrary DM particles, and $h$ the Higgs boson. This process presents an average cross section times relative velocity given by
\begin{eqnarray}\label{a}
 \braket{\sigma_{\psi'h}v} = \frac{16\pi}{J}m_\psi^2 \Phi_{h},\quad\quad\text{with}\quad\quad E_h = m_\psi\left(1 - \frac{m_{\psi '}^2 - m_h^2}{4m_\psi^2}\right)
\end{eqnarray}
with $J$ an arbitrary $J$-factor, $m_\psi$ the mass of the initial states, $\Phi_h$ is the flux, and the outgoing Higgs having an energy $E_h$ (equivalently to \ref{energies}). Now, it is possible to have the same flux $\Phi_h$ with an outgoing Higgs having the same energy $E_h$ in another process given by $\psi\bar{\psi}\rightarrow hh$:
\begin{eqnarray}\label{b}
 \Phi_h = \frac{1}{2}\frac{\braket{\sigma_{hh} v}}{8\pi E_h^2}J, 
\end{eqnarray}
where the 1/2 factor comes from the fact that we are considering one outgoing Higgs, and we have set $m_\psi = E_h$. Combining \ref{a} and \ref{b} we obtain
\begin{figure}[t!]
\centering
\includegraphics[width=0.35\textwidth]{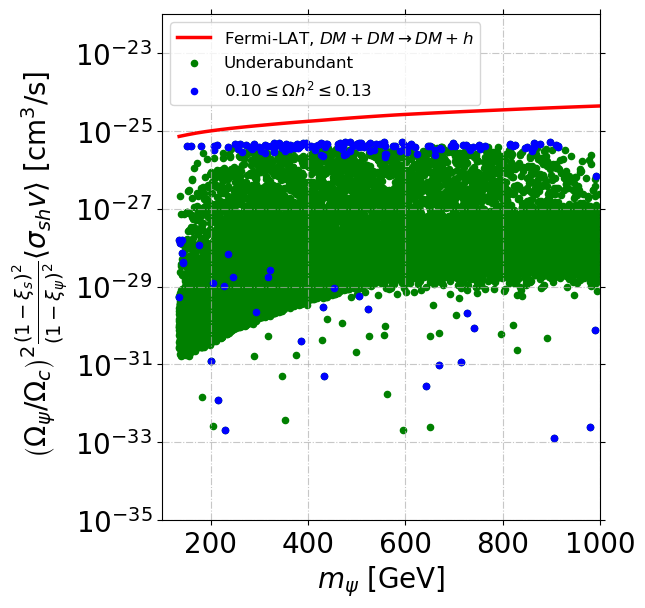}\quad\quad\quad
\includegraphics[width=0.35\textwidth]{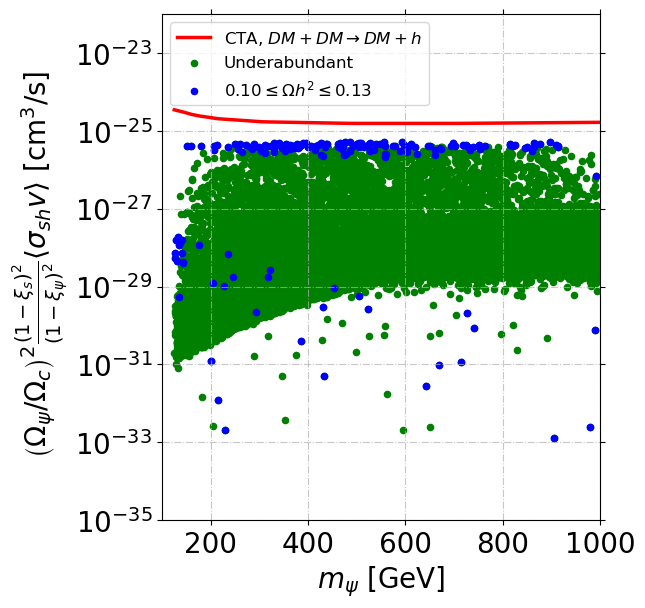}\quad\quad\quad
\includegraphics[width=0.35\textwidth]{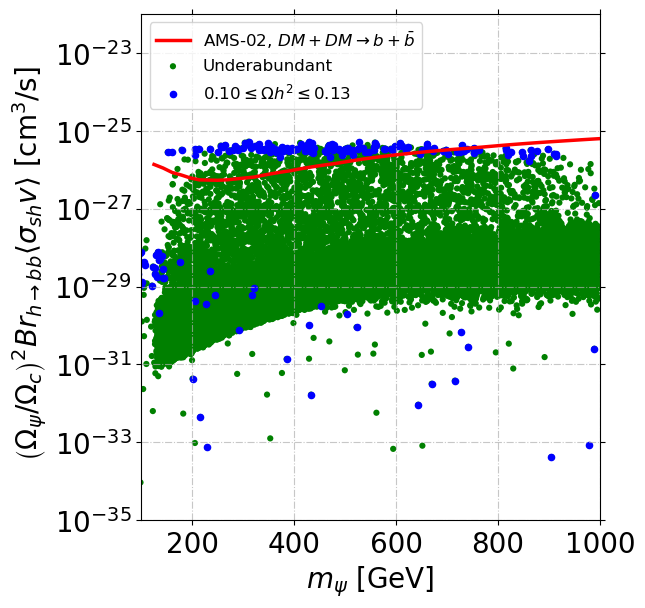}\quad\quad\quad
\includegraphics[width=0.35\textwidth]{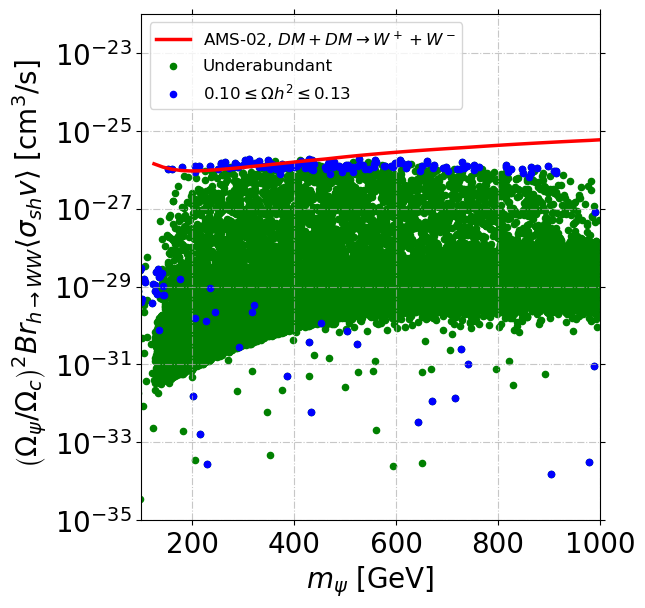}
\caption{\textit{Points with $\Delta > 0$ fulfilling direct detection constraints as a function of the re-scaled average annihilation cross section times velocity given by expressions \ref{gammatrans} and \ref{f} multiplied by the relative abundance $(\Omega_\psi/\Omega_c)^2$ of the annihilating fermion DM. Upper bounds for Fermi-LAT \cite{Queiroz:2019acr}, CTA \cite{Queiroz:2019acr} and AMS-02 \cite{Reinert:2017aga} are given by the red continuous lines. For details see the main text.}}
\label{upperid}
\end{figure}
\begin{eqnarray}\label{c}
 \braket{\sigma_{\psi'h}v} = \frac{1}{(1 - \xi_{\psi'})^2}\braket{\sigma_{hh}v},
\end{eqnarray}
with $\xi_x \equiv \left(m_x^2 - m_h^2\right)/(4m_\psi^2)$, with $x$ being some particle in the final state. From the last relation $\psi'$ is an arbitrary state, then in eq.~\ref{c} taking $\psi' = \psi$ and $\psi' = s$ and combining the resulting expressions, it follows that
\begin{eqnarray}\label{gammatrans}
\left(\frac{1 - \xi_s}{1 - \xi_\psi}\right)^2\braket{\sigma_{s h} v}=  \braket{\sigma_{\psi h} v},
\end{eqnarray}
Upper limits on $\braket{\sigma_{\psi h}v}$ \cite{Queiroz:2019acr}, now can be compared to our $\braket{\sigma_{s h} v}$ for certain values of $m_\psi$ and $m_s$ through eq.~\ref{gammatrans}. In Fig.~\ref{upperid}~(\textit{top row}) we show the projection of under-abundant points (green) and those that give approximately the correct relic abundance (blue points) for the resulting average annihilation cross section times velocity (left side of eq.~\ref{gammatrans}) times the relative abundance of the annihilating DM particles. The red line in both plots corresponds to the upper bound found in \cite{Queiroz:2019acr} (right side of eq.~\ref{gammatrans}). All these points are in the region $\Delta > 0$ and fulfill direct detection constraints. None of the bounds touch the parameter space points, then not showing any possible exclusion. Note that the blue points well below the upper bounds (red line) correspond to the Higgs resonance region, since lower couplings $(g_\psi,\lambda_{hs})$ are necessary to compensate the resonance effect, then reducing the ID signals.

On the other hand, AMS-02 anti-protons measurement set upper bounds on $\braket{\sigma_{\text{DM DM}\rightarrow X\bar{X}} v} \equiv \braket{\sigma_{X\bar{X}} v}$ with $X\bar{X} = b\bar{b}, W^+W^-$ \cite{Reinert:2017aga}. From the semi-annihilation with the Higgs decaying into $b\bar{b}$ we have that 
\begin{eqnarray}\label{d}
 \text{Br}(h\rightarrow b\bar{b})\braket{\sigma_{s h} v} = \frac{8\pi}{J}m_\psi^2 \Phi_{b\bar{b}}
\end{eqnarray}
Additionally, the flux $\Phi_{b\bar{b}}$ in \ref{d} can be produced by an arbitrary interaction $\psi\bar{\psi}\rightarrow b\bar{b}$: 
\begin{eqnarray}\label{e}
 \Phi_{b\bar{b}} = \frac{\braket{\sigma_{b\bar{b}} v}}{8\pi m_\psi^2}J
\end{eqnarray}
Combining \ref{d} and \ref{e} we obtain 
\begin{eqnarray}\label{f}
 \text{Br}(h\rightarrow b\bar{b})\braket{\sigma_{s h} v} = \braket{\sigma_{b\bar{b}} v}
\end{eqnarray}
Equivalently for $W^+W^-$. In Fig.~\ref{flux_micro}~(\textit{bottom row}) we show the projection of points considering the semi-annihilation with Higgs decay into $b\bar{b}$ and $W^+W^-$, contrasted with the upper bounds given by AMS-02. From the resulting plots, $b\bar{b}$ search set the stringent bounds on the parameter space of the model, discarding points with the correct relic abundance with masses below $\sim 500 - 700$ GeV, approximately.

\section{Discussion and Conclusions}
In this paper we have explored a simple extension to the SM containing two gauge-singlet fields: a Dirac fermion and a real pseudoscalar. From the DM point of view, the model present multiple scenarios with one or two DM components. Previous works have focused on the scenario when the Higgs portal coupling takes very small values, in such a way that the dark sector evolves decoupled from the SM, with the DM relic abundance produced via dark freeze-out/freeze-in. In the present work we found that as the Higgs portal take sizable values it is possible to produce a one-component DM candidate via freeze-in or a two-component DM via freeze-out. We have focused on the latter scenario, which only presents four free parameters: two masses and two couplings.

The stability of the two singlets is guaranteed by parity arguments without the necessity of invoking additional symmetries. Furthermore, the relic abundance of both singlets is determined via freeze-out through annihilations, DM conversions and semi-annihilations. The appearance of the latter is contrary to the standard belief that this type of processes appear only in the presence of symmetries larger than $Z_2$. We have explored the relic density abundance of both DM components in the parameter space, founding interesting behaviors depending on the mass hierarchy and couplings values. Semi-annihilations and DM conversions play an important role in two of the three available regions, making the pseudoscalar relic abundance low enough to evade direct detection XENON1T bounds, and then allowing DM with masses of hundreds of GeV up to the TeV scale. As it is usual with Higgs portals, the resonance region will not be discarded completely, even with the powerful projection of XENONnT. 

We have complemented our analysis with indirect detection signatures and bounds. Firstly, we have explored the box-shape gamma rays signals appearing from a fermion DM semi-annihilation, showing small boxes signals for the fermion DM $s$-channel semi-annihilation. Furthermore, we have translated bounds from gamma-ray searches from Fermi-LAT and CTA projections onto the semi-annihilation, not showing any possible tension, although the sensitivity of both experiments rely in the ballpark of part of the parameter space of the model. In fact, considering that our numerical methods are not very precise, a more detailed and sophisticated analysis such as a global-fit could be perfectly sensitive to these gamma-rays upper bounds in view of the expected quantitative differences between the two methods. As a final analysis, we have tested the model with anti-protons upper bounds from AMS-02 experiment, showing an exclusion of fermion DM masses below $\sim 500 - 700$ GeV. 

Finally, considering that the viability of the model requires high values for the dark sector coupling $g_\psi$ in some regions of the parameter space, there are two important points to be taken into account, although the precise calculation of them are beyond the scope of the present work. The first is related to the possible appearance of Landau poles at low energy scales, implying an urgent UV completion of the model (e.g. embedding the pseudoscalar into a complex singlet which acquires vev). The second point is related to the spin-independent one-loop amplitude in direct detection for the singlet fermion, which due to the high coupling values it may give a sizable contribution, possibly excluding even more parameter space of the model, in particular the special region in which the relic abundance of the pseudoscalar drops to zero.


\section*{Acknowledgments}
We would like to thank Alejandro Ibarra, Camilo García, Alexander Belyaev and Claudio Dib for useful discussions, and to Alexander Pukhov for helping with MicrOMEGAS. B.D.S would like also to ANID (ex CONICYT) Grant No. 74200120. A.Z. has been partially founded by ANID (Chile) PIA/APOYO AFB 180002, and P. E. has been partially founded by project FONDECYT N° 1170171.


\appendix
\section{Annihilation Cross Sections}\label{app}
The exact values of $\braket{\sigma v}$ for the different $2\rightarrow 2$ processes were calculated with \texttt{micrOMEGAs 5.2.7a}. In order to show the dependence of the thermally average cross section on the parameters of the model, in this appendix we show some expressions for $\braket{\sigma v}$ in the limit in which $\braket{\sigma v} \approx (\sigma v)|_{s = (m_1 + m_2) (1 + v^2/4)}$, with $m_1$ and $m_2$ the masses of the annihilating particles, keeping the $s$-wave only when it is present:
\small
\begin{eqnarray}\label{sigmav1}
 \braket{\sigma v}_{\psi + \bar{\psi}\rightarrow s+h} &=& \frac{g_\psi^2\lambda_{hs}^2v_H^2\sqrt{-2m_h^2(m_s^2 + 4m_\psi^2) + m_h^4 + (m_s^2 -4m_\psi^2)^2}}{64\pi m_\psi^2(4m_\psi^2 - m_s^2)^2}, \\
                                   \braket{\sigma v}_{\psi + \bar{\psi}\rightarrow s + s}  &=& \frac{g_\psi^4 m_\psi (m_\psi^2 - m_s^2)^{5/2}}{24\pi(m_s^2 - 2m_\psi^2)^4}  v^2 \\
                                   \braket{\sigma v}_{s + s\rightarrow \psi + \bar{\psi}}  &=& \frac{g_\psi^4 (m_s^2 - m_\psi^2)^{3/2}(2m_s^2 + 3m_\psi^2)}{60\pi m_s^7}  v^4 \\
               \braket{\sigma v}_{\psi + s\rightarrow \psi + h}   &=& -\frac{g_\psi^2\lambda_{hs}^2v_H^2\sqrt{(m_s^2 - m_h^2)((m_s + 2m_\psi)^2 - m_h^2)}\left(2m_\psi - \frac{-m_h^2 + m_s^2 + 2m_sm_\psi + 2m_\psi^2}{m_s + m_\psi}\right)}{32\pi m_s(m_s + m_\psi)^2\left(m_\psi\frac{-m_h^2 + m_s^2 + 2m_sm_\psi + 2m_\psi^2}{m_s + m_\psi} + m_s^2 - 2m_\psi^2\right)^2}
\end{eqnarray}
\normalsize
The corresponding expressions for the annihilations of a pair of $s$ into SM particles can be found in the appendix of \cite{Cline:2013gha}.

\bibliography{bibliography}
\bibliographystyle{utphys}

\end{document}